\documentclass[prd,notitlepage,longbibliography,nofootinbib,superscriptaddress,twocolumn,preprintnumbers]{revtex4-2}

\usepackage[utf8]{inputenc}

\usepackage{bm}
\usepackage{comment} 
\usepackage[colorlinks=true,urlcolor=blue,anchorcolor=black,citecolor=blue,linkcolor=black,filecolor=black,menucolor=black,pagecolor=black,linktocpage=true,pdfproducer=medialab,pdfa=true]{hyperref}
\usepackage{graphicx}
\usepackage{amsmath,latexsym,amssymb,mathrsfs,ascmac}
\usepackage{multirow}
\usepackage{braket}

\newcommand{\nn}{\nonumber \\}

\begin{document}
\preprint{UT-Komaba/24-1}

\title{Gedanken Experiments to Destroy a Black Hole by a Test Particle:
\\[2mm]
Multiply Charged Black Hole with Higher Derivative Corrections}

\author{Keisuke Izumi}
\affiliation{Kobayashi-Maskawa Institute, Nagoya University, Nagoya 464-8602, Japan}
\affiliation{Department of Mathematics, Nagoya University, Nagoya 464-8602, Japan}

\author{Toshifumi Noumi}
\affiliation{Graduate School of Arts and Sciences, The University of Tokyo, Komaba, Meguro-ku, Tokyo 153-8902, Japan}

\author{Daisuke Yoshida}
\affiliation{Department of Mathematics, Nagoya University, Nagoya 464-8602, Japan}

\begin{abstract}

We investigate a gedanken experiment to destroy an extremally charged black hole by dropping a test particle, provided that there are multiple $U(1)$ gauge fields coupled with each other through higher derivative interactions.
In the absence of higher derivative corrections, it is known that the Coulomb repulsion prevents a test particle that would break the extremal condition from
falling into an extremal black hole and therefore the black hole cannot be destroyed.
We extend this observation to include higher derivative corrections.
Although the extremal condition is modified by the higher derivative interactions, we find that the repulsive force induced by the higher derivative couplings is responsible for preventing a test particle that would break the modified extremal condition to reach the event horizon.
Thus, we confirm that the weak cosmic censorship conjecture holds for extremally charged black holes even in the presence of higher derivative corrections, as long as the test particle approximation is justified. 

\end{abstract}
\maketitle

\section{Introduction}
The weak cosmic censorship conjecture \cite{Penrose:1969pc} states that all of the spacetime singularities formed by the gravitational collapse must be hidden by event horizons. This conjecture is supported by many gedanken experiments. In particular, gedanken experiments trying to destroy a black hole by dropping a test particle have been actively investigated so far. The pioneering work along this direction is Ref.~\cite{Wald:1974hkz}, where the motion of the test particle on the extremal Kerr--Newman spacetime is investigated. There, it was found that a test particle that would overcharge and/or overspun the extremal Kerr--Newman black hole is prevented to reach the event horizon due to the Coulomb repulsion, centrifugal force, and spin-spin interactions. Thus, such a test particle cannot destroy the black hole. Following this work, the analysis is extended to other kinds of black holes \cite{Cohen:1979zzb,Needham:1980fb,Semiz:1990fm,Bekenstein:1994nx,Semiz:2005gs}. In this paper, we investigate gedanken experiments along the same direction but with including the effective field theory corrections to Einstein gravity with two gauge fields.

If we regard the classical gravitational theory as a low energy effective field theory, it is natural to expect the presence of higher derivative corrections. For example, starting from the quantum electrodynamics with gravity, 
massive heavy charged fields generate the higher derivative corrections in the low energy effective action below their mass scale~\cite{Drummond:1979pp}.
Implications of such correction terms in the low energy effective theory are key bypaths to reveal the nature of the high energy physics. 
With the presence of higher derivative corrections, 
the extremal condition of black holes is modified (see, e.g., \cite{Kats:2006xp, Cheung:2018cwt, Hamada:2018dde, Jones:2019nev}), 
because the electro-vacuum solutions have corrections from the Reissner--Nordstr\"{o}m solution. 
This modification would have the information of the high energy physics.

In this paper, we investigate whether such modified extremal conditions could be broken by a test particle. 
We treat an extremally charged black hole in the theory of Einstein gravity with two $U(1)$ gauge fields including higher derivative corrections.
We use the test particle approximation, which is valid if the backreaction to the test particle by itself is negligible. 
Hence, this approximation is trustable in the first order in the charge of the test particle. 
Within this approximation, we confirm that any charged test particle can never destroy an extremal black hole, that is, 
the weak cosmic censorship conjecture holds.

If the charges of the black hole and the particle are associated with different gauge fields, no Coulomb force acts between them. 
However, it implies that the first order in the charge of the test particle vanishes, and 
the higher order contributions are required to be taken into account. 
The importance of effects beyond the test particle approximation, such as the self-force or self-energy effects \cite{Leaute:1982sm}, are discussed in Refs.~\cite{Hod:2002pm, Sorce:2017dst}. 
In particular, the Sorce--Wald formalism~\cite{Sorce:2017dst}, which does not rely on the test point particle approximation, enables us to include such effects as the second order increase of the mass of the resultant black hole. 
The second order effects for the sub-extremal Kerr--Newman black hole is investigated in the original paper~\cite{Sorce:2017dst}, and, then, the formalism and its higher order extension are studied in Refs.~\cite{Jiang:2019soz, Wang:2020vpn, Duztas:2021kni, Sang:2021xqj, Wang:2022umx, Chen:2020hjm,Jiang:2021ohh,Lin:2022ndf}.
We note also that the gedanken experiments to destroy the extremal black hole with higher derivative corrections with the single gauge field is investigated in Refs.~\cite{Chen:2020hjm,Jiang:2021ohh} based on the Sorce--Wald formalism at the first order in a charge. See also Ref.~\cite{Lin:2022ndf} for the higher order analysis for the sub-extremal black hole with higher derivative corrections. 
The higher order contributions in the particle charges are not addressed in this paper, but the similar results are expected to be obtained.

\begin{table*}[htb]
\begin{tabular}{c|c|c|c|c}
gauge field & field strength& electrostatic potential& particle charge & black hole charge
\\
\hline
 $A_{\mu}$ & $F_{\mu\nu}$ & $\Phi$ & $q$ & $Q$ \\
 $B_{\mu}$ & $H_{\mu\nu}$ & $\Psi$ & $p$ & $P$ \\
\end{tabular}
\caption{Summary of the Notation}
\label{tbl.1}
\end{table*}

This paper is organized as follows. 
In the next section, we show the dynamics of a test charged particle on the static, spherically symmetric spacetime and electric fields. 
Then, we give the gedanken experiments for the black hole with minimally coupled gauge fields in the section \ref{sec.3}. 
In the section \ref{sec.4}, we generalize the analysis to the case with higher derivative corrections. 
The final section is devoted to the summary and discussion. 
We describe the detailed equations in the appendix \ref{sec.A}. 
In the appendix \ref{sec.B}, we confirm that the results do not change under the field redefinition.
Throughout the paper, we represents two $U(1)$ gauge fields by $A_{\mu}$ and $B_{\mu}$. The field strength, the charge of the particle, and the charge of the black hole with respect to each gauge fields are summarized in the table \ref{tbl.1}. We use the unit $\hbar = 1, c = 1$, and $\epsilon_{0} = 1$, where $\hbar, c$ and $\epsilon_{0}$ are the reduced Planck constant, the speed of light, and the permittivity of vacuum.

\section{Motion of Test Particle with Multiple Charges}
\label{sec.2}
In this section we summarize the dynamics of a test charged particle on the static and spherically symmetric spacetime with two $U(1)$ gauge fields $A_{\mu}$ and $B_{\mu}$.
The particle that we consider has mass $m$ and charges $q$ and $p$ with respect to the gauge fields $A_{\mu}$ and $B_{\mu}$, respectively. 
Suppose that the particle is minimally coupled with gravity and have no derivative coupling with the background fields. 
This setup is justified in our analysis where the lowest order corrections with higher derivative couplings are taken into account, 
since they are removed by the field redefinition.%
\footnote{The direct analysis with keeping the lowest order corrections is shown in Appendix~\ref{AppFC}.}
The action of the test particle is
\begin{align}
 S^{\text{test}}[x^{\mu}] = \int d\tau \biggl[& - m \sqrt{- g_{\mu\nu}(x) \dot{x}^{\mu} \dot{x}^{\nu}} \notag\\
& + q A_{\mu}(x) \dot{x}^{\mu}
+ p B_{\mu}(x) \dot{x}^{\mu}
 \biggr],
 \label{Actionparticle}
\end{align}
where $x^{\mu}(\tau)$ is the worldline of the test particle and the dot represents the derivative with respect to $\tau$.
This action has the symmetry under the reparametrization of $\tau$. 
We fix this degree of freedom by setting $\tau$ to be the proper time, {\it i.e.}, $g_{\mu\nu} \dot{x}^{\mu} \dot{x}^{\nu} = - 1$. 
Then, the equation of motion for the test particle reads
\begin{align}
m u^{\nu} \nabla_{\nu} u^{\mu} &= q F^{\mu}{}_{\nu} u^{\nu} + p H^{\mu}{}_{\nu} u^{\nu}, \\
u_{\mu} u^{\mu} &= -1,
\end{align}
where $u^{\mu}$ is the four-velocity of the particle, that is, $u^{\mu} := \dot{x}^{\mu}$.
Here, $F_{\mu\nu}$ and $H_{\mu\nu}$ are the field strength of $A_{\mu}$ and $B_{\mu}$, defined by $F_{\mu\nu} := \partial_{\mu} A_{\nu} - \partial_{\nu} A_{\mu}$ and $H_{\mu\nu} := \partial_{\mu} B_{\nu} - \partial_{\nu} B_{\mu}$.

Let us focus on the static spherically symmetric case, and denote the static Killing vector by $\xi^{\mu}$. 
Then, we introduce the static time coordinate $t$ by $\xi^{\mu}\partial_{\mu} = \partial_{t}$. 
With the static time $t$ and the areal radius $r$, the static spherically symmetric metric can be described as
\begin{align}
g_{\mu\nu} dx^{\mu} dx^{\nu} = - f(r) dt^2 + \frac{h(r)}{f(r)}dr^2 + r^2 d\Omega^2,\label{ansatzg}
\end{align} 
where $d\Omega^2 := d\theta^2 + \sin^2 \theta d\phi^2$ is the metric on the unit two sphere. 
In asymptotically flat spacetime, the event horizon $r_{\text{H}}$ is located at the largest positive root of $f(r)$, if it exists. 
We focus only on the outside of the black hole, that is, the region $r \in (r_{\text{H}},  \infty)$, where $f(r)$ and $h(r)$ are assumed to be positive.
The asymptotic flatness implies $f(r) \rightarrow 1$ and $h(r) \rightarrow 1$ in the limit $r \rightarrow \infty$.

Suppose that $U(1)$ electric fields $A_\mu$ and $B_\mu$ also enjoy the static, spherically symmetric properties, and then, they take the following form of vector potentials:
\begin{align}
 A_{\mu} dx^{\mu} &= - \Phi(r) dt,\label{ansatzA} \\
 B_{\mu} dx^{\mu} &= - \Psi(r) dt.\label{ansatzB}
\end{align}
Here, $\Phi$ and $\Psi$ are the electrostatic potentials for $A_{\mu}$ and $B_{\mu}$, respectively. 
Note that they are manifestly static, that is, $(\mathsterling_{\xi} A)_{\mu} = 0$ and $(\mathsterling_{\xi} B)_{\mu} = 0$.
We also assume that the electrostatic potentials vanish at asymptotic infinity, {\it i.e.}, $\Phi(r) \rightarrow 0$ and $\Psi(r) \rightarrow 0$ in the limit $r \rightarrow \infty$. 
The corresponding field strengths are
\begin{align}
 \frac{1}{2} F_{\mu\nu} dx^{\mu} \wedge dx^{\nu} &= \Phi'(r) dt \wedge dr, \\
 \frac{1}{2} H_{\mu\nu} dx^{\mu} \wedge dx^{\nu} &= \Psi'(r) dt \wedge dr,
\end{align}
where the prime represents the derivative with respect to $r$.

Due to the static symmetry, the energy of the charged particle defined by
\begin{align}
 E := \left( - m g_{\mu\nu} u^{\nu} - q A_{\mu} - p B_{\mu} \right) \xi^{\mu},
\end{align}
is conserved along the worldline, which implies $u^{\mu} \nabla_{\mu} E = 0$.

Let us investigate a test particle moving along the radial direction, starting from the asymptotic infinity, that is, 
$r(\tau) \rightarrow \infty$ at the infinite past $\tau \rightarrow - \infty$. 
We denote the four-velocity $u^{\mu}$ by
\begin{align}
 u^{\mu} \partial_{\mu} = \dot{t}(\tau) \partial_{t} + \dot{r}(\tau) \partial_{r},
\end{align}
with $\dot{t} > 0$ and $\dot{r} < 0$.
Then, the proper time condition and the conserved energy can be written as,
\begin{align}
 -1 &= - f(r) \dot{t}^2 + \frac{h(r)}{f(r)} \dot{r}^2,  \\
 E &= m f(r) \dot{t} + q \Phi(r) + p \Psi(r).
\end{align}
By eliminating $\dot{t}$ from the above equations, we obtain
\begin{align}
 E = m \sqrt{f + h \dot{r}^2} + q \Phi + p \Psi.\label{E=}
\end{align}
Due to the energy conservation, the value of $E$ can be fixed in the asymptotic region as
\begin{align}
 \lim_{r \rightarrow \infty} E = m \sqrt{1 + \dot{r}^2},
\end{align}
and hence we focus on the case $E \geq m$.
By solving Eq.~\eqref{E=} for $\dot{r}^2$, we obtain
\begin{align}
 \dot{r}^2  &= \frac{1}{m^2 h(r)}\left( (q \Phi + p \Psi - E)^2 - m^2 f(r) \right), \notag\\  
 &= \frac{1}{m^2 h(r)}\left(E -  V_{+}(r)\right) \left( E - V_{-}(r) \right),
\end{align}
where $V_{\pm}$ are defined by
\begin{align}
V_{\pm}(r) := q \Phi(r) + p \Psi(r) \pm m \sqrt{f(r)}.
\label{Vpm}
\end{align}
From Eqs.~\eqref{E=} and \eqref{Vpm}, we find that the positivity of $\dot{r}^2$ requires 
\begin{align}
 E \geq V_{+}(r).\label{EgeqVp}
\end{align}
This inequality indicates the allowed region of motion for any given energy $E \geq m$. 
Thus, $V_{+}(r)$ can be regarded as the effective potential of this system. In particular, we obtain a necessary condition for the particle to reach the event horizon $r_{\text{H}}$, which we call the {\it falling condition}, by
\begin{align}
 E \geq V_{+}(r_{\text{H}}) = q \Phi(r_{\text{H}}) + p \Psi(r_{\text{H}}),\label{falloff}
\end{align}
where $V_{+}(r_{\text{H}})$ can be understood as the $r \rightarrow r_{\text{H}}$ limit of $V(r)$ and so on.
This can be regarded as the condition among $E,p$ and $q$ for the test particle falling into the black hole.
In the following section, we will compare this condition with the extremal condition of the total system.

\section{Black Hole with Two Minimally Coupled Gauge Fields}
\label{sec.3}
In this section, we investigate the gedanken experiments to destroy a black hole in the theory of Einstein gravity with two minimally coupled $U(1)$ gauge fields without higher derivative interactions between the gauge fields.

\subsection{Static Spherically Symmetric Solution}

We consider the Einstein--Hilbert action with two minimally coupled $U(1)$ gauge fields, 
\begin{align}
\label{action_minimal}
 S = \int d^4 x \sqrt{- g} \left[
\frac{1}{16 \pi G} R - \frac{1}{4} F_{\mu\nu} F^{\mu\nu} - \frac{1}{4} H_{\mu\nu} H^{\mu\nu}
\right].
\end{align}
The Einstein equation is
\begin{align}
 R_{\mu\nu} - \frac{1}{2} g_{\mu\nu} R = 8 \pi G T_{\mu\nu},
\end{align}
where $T_{\mu\nu}$ is given by
\begin{align}
 T_{\mu\nu} &= F_{\mu\rho} F_{\nu}{}^{\rho} - \frac{1}{4} g_{\mu\nu} F_{\rho\sigma} F^{\rho\sigma} \notag\\
& \qquad +   H_{\mu\rho} H_{\nu}{}^{\rho} - \frac{1}{4} g_{\mu\nu} H_{\rho\sigma} H^{\rho\sigma}.
\label{EMTL}
\end{align}
The Maxwell equations for each gauge field are
\begin{align}
 \nabla_{\nu} F^{\nu \mu} &= 0, \\
 \nabla_{\nu} H^{\nu \mu} &= 0.
\end{align}
By solving the above equations with the static, spherically symmetric ansatz \eqref{ansatzg}, \eqref{ansatzA}, and \eqref{ansatzB}, we obtain the Reissner--Nordstr\"{o}m solution as usual,
\begin{align}
f(r) &= 1 - \frac{2 G M}{r} + \frac{G \left(Q^2 + P^{2}\right)}{4 \pi r^2}, \\
h(r) &= 1,\\
 \Phi(r) &= \frac{Q}{4 \pi r},\\
 \Psi(r) &= \frac{P}{4 \pi r} .
\end{align}

The roots of $f(r) = 0$ are 
\begin{align}
 r = G \left( M \pm \sqrt{M^2 - \frac{Q^2 + P^2}{4 \pi G}} \right).
\end{align}
Then, in terms of the extremal mass $\bar{M}_{\text{ext}}(Q,P)$ defined by\footnote{The bar of $\bar{M}_{\text{ext}}(Q,P)$ indicates that the extremal mass for given $Q$ and $P$ are evaluated in the minimally coupled theory~\eqref{action_minimal}. We will use similar notation for other quantities as well.}
\begin{align}
\label{def_M_bar_ext}
 \bar{M}_{\text{ext}}(Q,P) := \sqrt{\frac{Q^2 + P^2}{4 \pi G}},
\end{align}
the condition for the event horizon to exist, or in other words, the sub-extremal condition, can be written as
\begin{align}
 M \geq \bar{M}_{\text{ext}}(Q, P).
\end{align}
The equality holds for the extremal black hole. 
The areal radius of the event horizon of the extremal black hole becomes
\begin{align}
 \bar{r}_{\text{H}} = G \bar{M}_{\text{ext}}(Q,P).
\end{align}
On the other hand, the condition for the absence of the event horizon, say, the super-extremal condition, can be written as
\begin{align}
 M < \bar{M}_{\text{ext}}(Q, P).
\end{align}

\subsection{Gedanken Experiments to Destroy an Extremal Black Hole: Single Minimally Coupled Gauge Field} 

Let us first review the gedanken experiments to destroy an extremally charged black hole with a single gauge field, 
which is the case with $P = 0$ and $p = 0$ in our analysis. 
This corresponds to the non-rotating version of the analysis on the Kerr-Newman spacetime in Ref.~\cite{Wald:1974hkz}.

Let us start with the extremal black hole with $P = 0$ as a background spacetime
\begin{align}
 (M, Q, P ) = (\bar{M}_{\text{ext}}(Q,0), Q, 0),
\end{align}
and introduce a charged test particle with $p = 0$
\begin{align}
 (E, q, p) = (E, q, 0),
\end{align}
where $\bar{M}_{\text{ext}}(Q,0)$ is simplified as
\begin{align}
\bar{M}_{\text{ext}}(Q, 0) = \frac{|Q|}{\sqrt{4 \pi G}}.
\end{align}
In the test particle approximation, $E \ll M$ and $|q| \ll |Q|$ are assumed.
The event horizon of the background spacetime is located at $\bar{r}_{\text{H}} = G \bar{M}_{\text{ext}}(Q,0)$.

Let us confirm the sub-extremal condition for the total system first.
The total energy $\widehat{M}$ and charges $\widehat{Q}, \widehat{P}$ in this system are
\begin{align}
(\widehat{M}, \widehat{Q}, \widehat{P}) = (\bar{M}_{\text{ext}}(Q,0) + E, Q + q, 0).
\end{align}
Suppose that this system is settled in a static spacetime and no energy escapes to infinity. 
The uniqueness theorem states that if the resultant spacetime is a black hole solution, on the one hand,
it must be the Reissner--Nordstr\"{o}m spacetime with $(\widehat{M}, \widehat{Q}, \widehat{P})$, which satisfies the sub-extremal condition. 
On the other hand, if the super-extremal condition 
\begin{align}
 \widehat{M} - \bar{M}_{\text{ext}}(\widehat{Q}, \widehat{P}) = E - \text{sgn}(Q) \frac{q}{\sqrt{4 \pi G}} + \mathcal{O}(q^2) < 0\label{supext1}
\end{align}
is satisfied, the resultant object cannot be a black hole, and thus, the black hole will be destroyed.

The falling condition is investigated as follows. 
Suppose that the particle will be captured by the black hole. 
Then, the inequality \eqref{falloff} must be satisfied, that is, the falling condition \eqref{falloff} reduces to
\begin{align}
 E \geq q \frac{Q}{4 \pi r_{\text{H}}} = \text{sgn}(Q) \frac{q}{\sqrt{4 \pi G}}.\label{falloff1}
\end{align}

The super-extremal condition \eqref{supext1} and the falling condition \eqref{falloff1} cannot be satisfied at the same time.  
This leads us to the conclusion that any particle which would destroy the extremal black hole cannot be captured by the black hole.  

\subsection{Gedanken Experiments to Destroy an Extremal Black Hole: Two Minimally Coupled Gauge Fields}
\label{sec.3c}

Let us move on to the discussion on general cases including $P \neq 0$ and $p \neq 0$.
The extremal condition is given by 
\begin{align}
 (M, Q, P ) = (\bar{M}_{\text{ext}}(Q,P), Q, P).
\end{align}
We introduce a test particle with charge $(q,p)$,
and assume $E \ll M$ and $|q|,|p| \ll \sqrt{Q^2+P^2}$.

The super-extremal condition for the total system is obtained in the following discussion.
The total energy $\widehat{M}$ and charges $\widehat{Q}$ and $\widehat{P}$ of this system are given by 
\begin{align}
(\widehat{M}, \widehat{Q}, \widehat{P}) = (\bar{M}_{\text{ext}}(Q) + E, Q+q, P + p).\label{MQPtot}
\end{align}
Assuming $|q|,|p| \ll \sqrt{Q^2+P^2}$, we can evaluate the super-extremal condition as
\begin{align}
 & \widehat{M} - \bar{M}_{\text{ext}}(\widehat{Q}, \widehat{P})
 \notag\\
&=  E - \frac{1}{\sqrt{4 \pi G}} \frac{q Q + p P}{\sqrt{Q^2 + P^2}} + {\cal O}(p^2, p q, q^2)  < 0.
\end{align}
This inequality is rewritten as
\begin{align}
 E < \frac{1}{\sqrt{4 \pi G}} \frac{q Q + p P}{\sqrt{Q^2 + P^2}}  + {\cal O}(p^2, p q, q^2). \label{supext2}
\end{align}

Then, let us investigate the falling condition~\eqref{falloff}.
It can be written as 
\begin{align}
 E \geq \frac{q Q + p P}{4 \pi \bar{r}_{\text{H}}} = \frac{1}{\sqrt{4 \pi G}} \frac{q Q + p P}{\sqrt{Q^2 + P^2}}. \label{falloff2}
\end{align}

The super-extremal condition \eqref{supext2} and the falling condition \eqref{falloff2} cannot be satisfied at the same time, 
which is the same result as that in the single gauge field case. 
Our purpose of this paper is to verify whether the same structure holds even if we include the higher derivative corrections. 
We investigate it in the next section.

Before closing this section, we would like to mention on the special parameter choice,
\begin{align}
 q Q + p P = 0, \label{zeroforce} 
\end{align}
which means that the charges of the black hole $(Q, P)$ are orthogonal to those of the particle $(q,p)$ and no Coulomb force acts between the black hole and the particle. Without loss of generality, let us focus on the case with $P = 0$ and $q = 0$.
Then, the super-extremal condition starts from the second order in $p^2$, 
\begin{align}
& \widehat{M} - \bar{M}_{\text{ext}}(\widehat{Q}, \widehat{P})  = E - \frac{1}{2} \frac{|Q|}{\sqrt{4 \pi G}} \frac{p^2}{Q^2} + {\cal O}(p^4) < 0.\label{mmq2.0}
\end{align}
Thus, a naive discussion may lead to the result that the condition for the event horizon to be destroyed can be written as
\begin{align}
 E < \frac{1}{2} \frac{|Q|}{\sqrt{4 \pi G}} \frac{p^2}{Q^2}.\label{supext2.0}
\end{align}
However, no Coulomb force seems to acts on the particle, so that the falling condition may reduce to
\begin{align}
 E \geq 0, \label{falloff2.0}
\end{align}
which is trivially satisfied for $E \geq m$. 
Thus, the conditions \eqref{supext2.0} and \eqref{falloff2.0}, as well as $E \geq m$, can be satisfied at the same time for a test particle with the energy
\begin{align}
\label{window}
m \leq  E <  \frac{1}{2} \frac{|Q|}{\sqrt{4 \pi G}}\frac{p^2}{Q^2}.
\end{align}
However, it is too early to conclude that this is a counterexample of the cosmic censorship conjecture:
Eq.~\eqref{mmq2.0} indicates that the potential violation of the extremal condition is at most the order $p^2$, 
whereas the test particle approximation is valid at the order $p^1$.
A careful analysis beyond the test particle approximation is therefore required. 
Indeed, a general formalism treating the effects beyond the test particle approximation was developed in Ref.~\cite{Sorce:2017dst} for {\it sub-extremal} Kerr--Newman black hole with a {\it single} gauge field,
showing that those sub-extremal black holes cannot be overcharged or overspun.
Its generalization to the extremal black hole will be necessary to make a conclusion if the cosmic censorship conjecture is satisfied or not in this setup.
We will not address this issue further in the present paper.
Instead, we will see that the weak cosmic censorship conjecture holds at the level of the test particle approximation, that is, at the order $p^1$, 
with the higher derivative corrections included.

\section{Higher Derivative Corrections}
\label{sec.4}
In this section, we extend the discussion of the gedanken experiments to the theory with the higher derivative corrections up to four derivatives.
We derive the super-extremal conditions (subSec. \ref{SECwHD}) and 
the falling conditions (subSec. \ref{sec.4b}) on black holes with two charges with the higher derivative corrections, 
and then, show that no overlap exists between the two.

\subsection{Super-Extremal Conditions}
\label{SECwHD}
Based on the spirit of the effective field theory approach, we treat all the possible higher derivative corrections up to four derivatives. More explicitly, we consider
\begin{widetext}
\begin{align}
 S = \int d^4 x \sqrt{- g} \Bigl(&
\frac{1}{16 \pi G} R - \frac{1}{4} F_{\mu\nu} F^{\mu\nu} - \frac{1}{4} H_{\mu\nu} H^{\mu\nu} \notag\\
&
+\alpha_{1} F_{\mu\nu} F^{\mu\nu} F_{\rho\sigma} F^{\rho\sigma}
+\alpha_{2} F_{\mu}{}^{\nu} F_{\nu}{}^{\rho} F_{\rho}{}^{\sigma} F_{\sigma}{}^{\mu}
+ \alpha_{3} R\, F_{\mu\nu} F^{\mu\nu}
+ \alpha_{4} R_{\mu\nu} F^{\mu\rho} F^{\nu}{}_{\rho} 
+ \alpha_{5} R_{\mu\nu\rho\sigma} F^{\mu\nu} F^{\rho\sigma}
\notag
\\ 
&
+ \beta_{1} F_{\mu\nu} F^{\mu\nu} F_{\rho\sigma} H^{\rho\sigma}
+\beta_{2} F_{\mu}{}^{\nu} F_{\nu}{}^{\rho} F_{\rho}{}^{\sigma} H_{\sigma}{}^{\mu}
+ \beta_{3} R\, F_{\mu\nu} H^{\mu\nu}
+ \beta_{4} R_{\mu\nu} F^{\mu\rho} H^{\nu}{}_{\rho}
+ \beta_{5} R_{\mu\nu\rho\sigma} F^{\mu\nu} H^{\rho\sigma} \notag\\
&
+ \gamma_{11} F_{\mu\nu} F^{\mu\nu} H_{\rho\sigma} H^{\rho\sigma}
+ \gamma_{12} F_{\mu\nu} H^{\mu\nu} F_{\rho\sigma} H^{\rho\sigma}
+ \gamma_{21} F_{\mu}{}^{\nu} F_{\nu}{}^{\rho} H_{\rho}{}^{\sigma} H_{\sigma}{}^{\mu}
+ \gamma_{22} F_{\mu}{}^{\nu} H_{\nu}{}^{\rho} F_{\rho}{}^{\sigma} H_{\sigma}{}^{\mu} \notag\\
& \qquad 
+ \gamma_{3} R\, H_{\mu\nu} H^{\mu\nu}
+ \gamma_{4} R_{\mu\nu} H^{\mu\rho} H^{\nu}{}_{\rho}
+ \gamma_{5} R_{\mu\nu\rho\sigma} H^{\mu\nu} H^{\rho\sigma}
\notag\\
&
+ \kappa_{1} F_{\mu\nu} H^{\mu\nu} H_{\rho\sigma} H^{\rho\sigma}
+ \kappa_{2} F_{\mu}{}^{\nu} H_{\nu}{}^{\rho} H_{\rho}{}^{\sigma} H_{\sigma}{}^{\mu} \notag \\
& + \zeta_{1} H_{\mu\nu} H^{\mu\nu} H_{\rho\sigma} H^{\rho\sigma}
+ \zeta_{2} H_{\mu}{}^{\nu} H_{\nu}{}^{\rho} H_{\rho}{}^{\sigma} H_{\sigma}{}^{\mu}
\Bigr).\label{actionhd}
\end{align}
\end{widetext}
Here, all the coefficients, $\alpha_{i}$, $\beta_{i}$, $\gamma_{i}$, $\kappa_{i}$ and $\zeta_{i}$, should be small enough, since the higher derivative terms are corrections to the first line in the right hand side of Eq.~\eqref{actionhd}.
We regard the orders of $\alpha_{i}$, $\beta_{i}$, $\gamma_{i}$, $\kappa_{i}$ and $\zeta_{i}$ as ${\cal O}(\epsilon)$ in a small constant $\epsilon$.
Moreover, in the action~\eqref{actionhd}, we omit the terms that are quadratic in the background equations of motion such as
\begin{align}
& \left(R_{\mu\nu} - \frac{1}{2} g_{\mu\nu} R - 8 \pi G T_{\mu\nu}\right)^2, \, R^2, \nn
& \nabla_{\mu} F^{\mu\nu} \nabla_{\rho} F^{\rho}{}_{\nu}, \,
\nabla_{\mu} F^{\mu\nu} \nabla_{\rho} H^{\rho}{}_{\nu}, \,
\nabla_{\mu} H^{\mu\nu} \nabla_{\rho} H^{\rho}{}_{\nu}.
\label{HDign}
\end{align}
The reason is as follows: First, the effective coupling constants in front of these operators are $\mathcal{O}(\epsilon)$. Second, variation of those terms is proportional to the background equations, which gives one more factor of $\mathcal{O}(\epsilon)$ when evaluated with the full solution. In total their contributions are $\mathcal{O}(\epsilon^2)$, which are higher orders ignored in our analysis.
Note that the first combination in \eqref{HDign} together with the second one was used to replace $R_{\mu\nu} R^{\mu\nu}$ term into some terms appearing in \eqref{actionhd}. 
Similarly, since the Gauss-Bonnet term, $R^2 - 4 R_{\mu\nu}R^{\mu\nu} + R_{\mu\nu\rho\sigma} R^{\mu\nu\rho\sigma}$, does not contribute to the equations of motion, $R_{\mu\nu\rho\sigma} R^{\mu\nu\rho\sigma}$ term can be rewritten as well. 
See the appendix \ref{sec.A1} for more detail. Also we emphasize that the deformation of the action by field redefinitions is not done on the action~\eqref{actionhd}. 
See the appendix \ref{sec.B} for more comments on the field redefinition.

Note that the higher derivative interactions generically appear in the low-energy effective action. 
For instance, $\alpha_{i}$ terms, as well as $\zeta_{1}, \zeta_{2}$ and $\gamma_{3}, \gamma_{4}, \gamma_{5}$, 
are noting but the Drummond--Hathrell effective action~\cite{Drummond:1979pp} 
which arise from the mediation of heavy massive particles with a single charge. 
Similarly, if there exist bi-charged heavy massive particles, $\beta_{i}$ terms and others are generated as the low-energy effective interactions.

The equations of motion can be derived as 
\begin{align}
& R_{\mu\nu} - \frac{1}{2} R g_{\mu\nu} - 8 \pi G T_{\mu\nu} = 8 \pi G \delta T_{\mu\nu}, \label{eomshd1} \\
& \nabla_{\nu} F^{\nu\mu} = \delta J_{F}^{\mu}, \label{eomshd2} \\
& \nabla_{\nu} H^{\nu\mu} = \delta J_{H}^{\mu}, \label{eomshd3}
\end{align}
where $\delta T_{\mu\nu}, \delta J_{F}^{\mu},$ and $\delta J_{H}^{\mu}$ represent the contribution from the higher derivative correction terms and the detailed expressions are given in the appendix~\ref{sec.A3}.

We solve a static spherically symmetric solution perturbatively by regarding the parameters $\alpha_{i}, \beta_{i}, \gamma_{i}, \kappa_{i}, \zeta_{i}$ are the first order in perturbations, say, ${\cal O}(\epsilon)$. Let us expand the dynamical variables as
\begin{align}
 f(r) = \bar{f}(r) + \delta f(r), \\
 h(r) = \bar{h}(r) + \delta h(r), \\
 \Phi(r) = \bar{\Phi}(r) + \delta \Phi(r), \\
 \Psi(r) = \bar{\Psi}(r) + \delta \Psi(r), 
\end{align}
where $\bar{f}, \bar{h}, \bar{\Phi}$ and $\bar{\Psi}$ are ${\cal O}(\epsilon^{0})$ and $\delta f, \delta h, \delta \Phi$ and $\delta \Psi$ are regarded as ${\cal O}(\epsilon)$. 
In the background, that is, ${\cal O}(\epsilon^{0})$, the equations of motion reduce to the usual Einstein--Maxwell equations, 
and hence we obtain Reissner--Nordstr\"{o}m solution. 
Then, in the ${\cal O}(\epsilon)$ equations of motion, we can plug the ${\cal O}(\epsilon^{0})$ solutions in the right hand side of Eqs,\eqref{eomshd1}--\eqref{eomshd3}, 
because they already involve ${\cal O}(\epsilon)$ coefficients such as $\alpha_{i}$.
By solving ${\cal O}(\epsilon)$ equations of motion, we obtain the following perturbative solutions:
\begin{widetext}
\begin{align}
  f(r) &= 1 - \frac{2 G M}{r} + \frac{G (Q^2  + P^2 )}{4 \pi r^2} 
- \frac{G}{\pi r^4}\left( Q^2(-2 \alpha_3 + \alpha_{5}) + Q P (-2 \beta_{3} + \beta_{5})
+ P^2 (-2 \gamma_{3} + \gamma_{5})
 \right)  \notag\\
& \qquad
+\frac{G^2 M}{\pi r^5}\left( Q^2 (-6 \alpha_{3} - \alpha_{4} + \alpha_{5}) + Q P (-6 \beta_{3}-\beta_{4}+\beta_{5})
+ P^2 (-6 \gamma_{3} - \gamma_{4} + \gamma_{5} )\right) \notag\\
& \qquad
- \frac{G}{80 \pi^3 r^6}
\biggl(Q^4 \tilde \alpha_{1}+ Q^3 P \tilde  \beta_{1} +Q^2 P^2 \tilde  \gamma_{1} + Q P^3 \tilde  \kappa_{1}+ P^4\tilde  \zeta_{1} \biggr) \nn
& \qquad
+\frac{G^3 \bar{M}_{\text{ext}}(Q,P)^2
}{5 \pi r^6}
\biggl( Q^2 \left( 20 \alpha_{3} + 4 \alpha_{4} - \alpha_{5} \right) 
+ Q P \left(  20 \beta_{3} + 4 \beta_{4} - \beta_{5} \right)
+P^2 \left( 20 \gamma_{3} + 4 \gamma_{4} - \gamma_{5}\right) \biggr)
+ {\cal O}(\epsilon^2),\label{solf} \\
 h(r) &= 1 + \frac{G}{\pi r^4}\left(  Q^2 \tilde \alpha_{3}  + Q P \tilde  \beta_{3} + P^2 \tilde  \gamma_{3}\right) + {\cal O}(\epsilon^2),\label{solh} \\
 \Phi(r) & = \frac{Q}{4\pi r}
+ \frac{G M (2 Q \alpha_{5} + P \beta_{5})}{2 \pi r^4}  - \frac{1}{{160 \pi^3 r^5}}
\biggl(
4Q^3 \tilde  \alpha_{1}+ 3Q^2 P \tilde  \beta_{1}  + 2 Q P^2 \tilde  \gamma_{1}+  P^3\tilde  \kappa_{1} \biggr) \nn
&\quad
- \frac{1}{{10 \pi r^5}}
\biggl( 
G^2 \bar{M}_{\text{ext}}(Q,P)^2
\Bigl( 2Q \bigl(  \alpha_{4} + 6  \alpha_{5} \bigr)
+P  \bigl( \beta_{4} + 6\beta_{5}  \bigr)
\Bigr)
-\frac{GQ}{4\pi} \Bigl( Q^2 \tilde \alpha_{3}  + Q P \tilde  \beta_{3} + P^2 \tilde  \gamma_{3}\Bigr)
\biggr)+ {\cal O}(\epsilon^2),\label{solphi}\\
 \Psi(r) &=
\frac{P}{4 \pi r} 
+ \frac{G M (2 P \gamma_{5} + Q \beta_{5})}{2 \pi r^4} 
- \frac{1}{160 \pi^3 r^5} \biggl(
Q^3  \tilde  \beta_{1} +2Q^2 P \tilde \gamma_{1}+ 3Q P^2 \tilde \kappa_{1} +4 P^3 \tilde \zeta_{1}
\biggr) \nn
& \qquad
- \frac{1}{10 \pi  r^5} \biggl(
G^2 \bar{M}_{\text{ext}}(Q,P)^2
\Bigl( Q ( \beta_{4} + 6\beta_{5})+ 2P(\gamma_{4} + 6 \gamma_{5})\Bigr)
-\frac{GP}{4\pi} \Bigl(
Q^2  \tilde\alpha_{3} +Q P \tilde \beta_{3} + P^2\tilde \gamma_{3} \Bigr)
\biggr)+ {\cal O}(\epsilon^2),\label{solpsi}
\end{align}
\end{widetext}
where
\begin{align}
&\tilde \alpha_1 := 2\alpha_1+ \alpha_2, \\
&\tilde \beta_1 := 2\beta_1+ \beta_2,\\
&\tilde \gamma_1 := 2\gamma_{11}+ \gamma_{21}+2\gamma_{12}+ \gamma_{22},\\
&\tilde \kappa_1 := 2\kappa_1+ \kappa_2,\\
&\tilde \zeta_1 := 2\zeta_1+ \zeta_2,
\end{align}
and 
\begin{align}
&\tilde \alpha_3 := 10\alpha_3+ 3\alpha_4+3\alpha_5,\\
&\tilde \beta_3 := 10\beta_3+ 3\beta_4+3\beta_5,\\
&\tilde \gamma_3 := 10\gamma_3+ 3\gamma_4+3\gamma_5,
\end{align}
and  $\bar{M}_{\text{ext}}(Q,P)$ is the the extremal mass without the higher derivative corrections defined in Eq.~\eqref{def_M_bar_ext}.

Now let us derive the extremal mass, i.e., the minimum black hole mass, for given charges $Q,P$. Recall that in the Einstein-Maxwell theory~\eqref{action_minimal} without higher derivative corrections, the extremal mass for given charges $Q,P$ was obtained by solving the two conditions
\begin{align}
\label{extremal_in_terms_of_f}
f(r_{\text{H}})=0\,,
\quad
f'(r_{\text{H}})=0
\end{align}
for $r_{\text{H}}$ and $M$. Here the first condition means that $r=r_{\text{H}}$ is a horizon, whereas the second condition is for the two horizons to degenerate. Since higher derivative corrections are supposed to be small, those properties of the metric functions should remain qualitatively the same and in particular the extremal mass should be determined by the conditions~\eqref{extremal_in_terms_of_f} as long as the derivative expansion works well. Indeed, one can explicitly confirm this expectation within the perturbative expansion in $\epsilon$.

Our task is then to solve Eq.~\eqref{extremal_in_terms_of_f} perturbatively. To do so, let us expand the horizon radius $r_{\text{H}}$ and the extremal mass $M_{\rm ext}(Q,P)$ around those of the Einstein-Maxwell theory without higher derivative corrections:
\begin{align}
r_{\text{H}}&=\bar{r}_{\text{H}}+\delta r_{\text{H}},
\\
M_{\rm ext}(Q,P)&=\bar{M}_{\rm ext}(Q,P)+\delta M_{\rm ext}(Q,P),
\end{align}
where as we defined earlier we have
\begin{align}
\bar{r}_{\text{H}}=G\bar{M}_{\rm ext}(Q,P)
,
\quad
\bar{M}_{\rm ext}(Q,P)=\sqrt{\frac{Q^2 + P^2}{4 \pi G}}.
\end{align}
\begin{widetext}
To solve the first condition of Eq.~\eqref{extremal_in_terms_of_f}, it is convenient to expand $f(r)$ as
\begin{align}
f(\bar r_{\text{H}}+ \delta r)&= - \frac{2 \delta M_{\rm ext}(Q,P)}{\bar{M}_{\text{ext}}(Q,P)}
- \frac{ G}{80 \pi^3 \bar r_{\text{H}}^6}  \Bigl( Q^4 \alpha + Q^3 P \beta +  Q^2 P^2 \gamma + Q P^3  \kappa + P^4  \zeta
\Bigr)
+ {\cal O}(\epsilon^2, \epsilon\,\delta r_{\text{H}}, \delta r_{\text{H}}^2) ,
\label{delf}
\end{align}
where 
the coefficients are defined by
\begin{align}
 \alpha &:= \tilde \alpha_1 + 4 \pi G (\alpha_{4}+\alpha_{5}), \\
 \beta &:=  \tilde  \beta_1 + 4 \pi G (\beta_{4}+\beta_{5}), \\
 \gamma &:= \tilde  \gamma_1 +4 \pi G (\alpha_{4} +  \gamma_{4}+\alpha_{5} +  \gamma_{5} ), \\
 \kappa &:= \tilde  \kappa_{1} + 4 \pi G  (\beta_{4}+\beta_{5}),  \\
 \zeta &:= \tilde  \zeta_{1}  + 4 \pi G  (\gamma_{4} +\gamma_{5} ).
\end{align}
Note that $\mathcal{O}(\delta r_{\text{H}})$ terms do not appear in the right hand side because of the extremal condition defining $\bar{r}_{\text{H}}$ terms. The correction to the extremal mass then reads
\begin{align}
\frac{\delta M_{\rm ext}(Q,P)}{\bar{M}_{\text{ext}}(Q,P)}=
 -\frac{2}{5} \frac{ G}{(4\pi)^3 \bar r_{\text{H}}^6}  \Bigl( Q^4 \alpha + Q^3 P \beta +  Q^2 P^2 \gamma + Q P^3  \kappa + P^4  \zeta
\Bigr)
+ \mathcal{O}(\epsilon^2) . \label{generalext}
\end{align}
Substituting this into the second condition $f'(r_{\text{H}})=0$ of Eq.~\eqref{extremal_in_terms_of_f} gives the shift $\delta r_{\text{H}}$ of the horizon position of the extremal black hole as
\begin{align}
 \frac{\delta r_{\text{H}}}{\bar{r}_{\text{H}}} = 
 - \frac{2G}{(4\pi)^{3} \bar r_{\text{H}}^6} \left( Q^4\tilde \alpha_1 +Q^3 P\tilde \beta_1 +Q^2P^2\tilde \gamma_1 +QP^3\tilde \kappa_1 +P^4\tilde \zeta_1 \right) 
+\frac{G}{\pi \bar r_{\text{H}}^4} \left( Q^2\alpha_3 +QP\beta_3 +P^2\gamma_3 \right)
 + {\cal O}(\epsilon^2). \label{correctedrH}
\end{align}
Here $\delta r_{\text{H}}$ is $\mathcal{O}(\epsilon)$ as expected.
Note that the areal radius of the event horizon of the extremal black hole is not invariant under the field redefinition. 
The reason is explained in appendix~\ref{AppA1a}.
\end{widetext}

\subsection{Falling Conditions}
\label{sec.4b}

Let us compare the super--extremal conditions with the falling conditions.
Our system is composed of an extremal black hole $M = M_{\text{ext}}(Q,P)$ and a test charged particle 
with the energy $E$ and charges $(q,p)$. The total mass and charges are 
\begin{align}
 (\widehat{M}, \widehat{Q}, \widehat{P}) = (M_{\text{ext}}(Q,P) + E, Q + q, P + p).
\end{align}
Let us expand the super-extremal condition for $(\widehat{M}, \widehat{Q}, \widehat{P})$ up to linear order in $q$ and $p$. 
The super-extremal condition $\widehat{M} < M_{\text{ext}}(\widehat{Q}, \widehat{P})$ can be expressed as 
\begin{align} 
E < E_\text{C}(Q,P;q,p)   + {\cal O}(\epsilon^2, q^2, q p, p^2),  \label{supext3}
\end{align}
with
\begin{widetext}
\begin{align} 
&E_\text{C}(Q,P;q,p) 
 \notag\\
& \quad
:= \frac{1}{4 \pi G \bar{M}_{\text{ext}}(Q,P)} \Biggl( 
 qQ +pP 
 \notag\\
& \qquad
+  \frac{2G}{5(4\pi)^3 r_{\text{H}}^6} \biggl( q \Bigl(
 Q^3(Q^2-4P^2) \alpha 
 +Q^2P(2Q^2-3P^2) \beta 
 +QP^2(3Q^2-2P^2)\gamma
 +P^3(4Q^2-P^2) \kappa  
 +5QP^4\zeta   \Bigr)
\notag\\
& \qquad \qquad 
+ p\Bigl(
 5Q^4P\alpha 
 +Q^3(-Q^2+4P^2)\beta  
 +Q^2P(-2Q^2+3P^2)\gamma 
 +QP^2(-3Q^2+2P^2) \kappa
+ P^3(-4Q^2+P^2)\zeta 
   \Bigr)
 \biggr) 
 \Biggr) . \notag\\
 \end{align}
\end{widetext}

Next, let us derive the falling condition. 
General expression has already been shown in Eq.~\eqref{falloff}.
Substituting the electrostatic potentials shown in Eq.~\eqref{solphi} and \eqref{solpsi} into \eqref{falloff}, we finally have
\begin{align}
 E &\geq q \Phi(r_{\text{H}}) + p \Psi(r_{\text{H}}) = E_\text{C}(Q,P;q,p),
 \label{falloff3}
\end{align} 
where $r_{\text{H}}$ is given by Eq.~\eqref{correctedrH}.
Thus, the super-extremal condition \eqref{supext3} and the falling condition \eqref{falloff3} cannot be satisfied at the same time at the leading order in $q$ and $p$. This is the main conclusion of this paper. 

To see the implications more explicitly, let us set $P = 0$ in the above equations. 
Then, the super-extremal condition $\widehat{M} < M_{\text{ext}}(\widehat{Q}, \widehat{P})$ can be expressed as 
\begin{align}
 E <
\frac{\text{sgn}(Q)}{\sqrt{4 \pi G}} 
\left( q
+ \frac{2 (\alpha q - \beta p )}{5 G^2 Q^2}
\right)
+ \mathcal{O}(\epsilon^2, p^2, pq,q^2),   \label{supext3P0}
\end{align} 
while the falling condition can be expressed as 
\begin{align}
 E &\geq  \frac{\text{sgn}(Q)}{\sqrt{4 \pi G}} \left(  q + \frac{2(\alpha q - \beta p)}{5G^2 Q^2}  \right)
+ {\cal O}(\epsilon^2).
\label{falloff3P0}
\end{align}
We would like to comment on specific choices of the particle charges.
First, if we set $p = 0$, our result reduces to the single gauge field case investigated in Ref.~\cite{Chen:2020hjm, Duztas:2021kni, Lin:2022ndf}. 

Next, let us consider the case with $q = 0$. 
Remember the discussion at the end of Sec.~\ref{sec.3} that, in the minimally coupled theory, this parameter choice corresponds to the case where
the ${\cal O}(p)$ contribution vanishes and the test particle approximation is violated. 
Contrary to the minimally coupled case, even though the condition $q = 0$ indicates no Coulomb force, there is additional repulsive forces 
that are induced through the higher derivative couplings between two gauge fields. 
More explicitly, from the expression \eqref{solpsi}, we can see that the electrostatic potential $\Psi$ is induced even when $P = 0$, through the coupling $\beta_{i}$. 
This is because even if objects (namely, particles and/or black holes) have different charges interact with each other through the mixing derivative couplings.
The condition for the absence of force at the event horizon can be written as 
\begin{align}
 q = \frac{2 \beta}{5 G^2 Q^2} p + {\cal O}(\epsilon^2) \label{noforceonhorizon}.
\end{align}
For the particle with this charges, 
both the Coulomb force 
and the force induced by the higher derivative corrections work but exactly cancel at the event horizon.
Then, ${\cal O}(q^2, qp, p^2)$ corrections are required to be seriously taken into account, 
because the effects beyond the point particle approximation might be significant. 
Thus the situation is similar to the case of $q = 0$ in the minimally coupled theory, although the particle feels Coulomb force at the infinity.

\section{Summary and Discussion}
\label{sec.5}
In this paper, we investigate gedanken experiments to destroy the extremal black hole by dropping a test charged particle. 
In particular, we focus on a theory with two $U(1)$ gauge fields with higher derivative corrections. 
The black hole we consider is a spherically symmetric, static extremal one with charges $(Q,P)$ with respect to two gauge fields. 
In the case with minimally coupled gauge fields, the black hole solution is described by Reissner--Nordstr\"{o}m solution. 
The super-extremal condition for the total energy and total charges of the system is given by Eq.~\eqref{supext2} and the falling condition, which shows that the particle can arrive at the event horizon, is given by Eq.~\eqref{falloff2}. 
Since these two conditions does not hold at the same time, extremal black holes cannot be destroyed by a test particle. 
We extend this analysis in the presence of the higher derivative corrections. 
Our action is given by Eq.~\eqref{actionhd}. 
The black hole solution is derived perturbatively in Eqs.~\eqref{solf} to \eqref{solpsi}. 
The super-extremal condition and the falling condition are obtained in Eq.~\eqref{supext3} and Eq.~\eqref{falloff3} respectively. 
As is the case of minimally coupled gauge fields, these conditions are not satisfied at the same time. 
Hence, we confirm that the extremal black hole with two gauge fields cannot be destroyed by a test particle 
even when the higher derivative corrections are included. 

Due to the validity of the test particle approximation, our result holds only up to the linear order in $q$ and $p$.
For a specific choice of the charges, Eq.~\eqref{zeroforce} for the minimally coupled case and Eq.~\eqref{noforceonhorizon} for the case with the higher derivative corrections with $P = 0$,
linear order contributions disappear both in the super-extremal condition and the falling condition. 
In such a case, the $O(p^2, p q, q^2)$ effects become dominant
 and the analysis beyond the point particle approximation similarly to the one in Ref.~Ref.~\cite{Sorce:2017dst}
is again required.

Although the second order analysis in Ref.~\cite{Sorce:2017dst} is applicable to sub-extremal black holes only and it does not directly apply to extremal black holes, it is instructive to see what kind of effects may be expected from a naive extremal limit of the result there: Here let us focus on the minimally coupled case for simplicity.
Assuming that the non-electromagnetic stress-energy tensor satisfies the null energy condition, Ref.~\cite{Sorce:2017dst} derived a lower bound on the second order correction to the total energy of the system, which captures the self-force and finite size effects for example. By extending the result there to the minimally coupled two gauge fields case naively, for a charged particle with $q=0$, $p\neq0$ thrown into a near-extremal black hole with $Q\neq0$, $P=0$, the lower bound reads
\begin{align}
\label{lower_b}
\widehat{M}\geq \bar{M}_{\text{ext}}(Q,0)+ \frac{1}{2} \frac{|Q|}{\sqrt{4 \pi G}} \frac{p^2}{Q^2} +\mathcal{O}(\varepsilon)\,,
\end{align}
where $\varepsilon$ is a small parameter quantifying deviation from the extremality of the original near-extremal black hole.
Here we assumed that the first order increase of the total energy is negligible.
Note that the second term on the right hand side is the lower bound on the second order correction to the total energy. On the other hand, the super-extremal condition on the total system up to $\mathcal{O}(p^2)$ is
\begin{align}
\label{extremality}
\widehat{M}-\bar{M}_{\text{ext}}(Q,p)=\widehat{M}-\bar{M}_{\text{ext}}(Q,0)- \frac{1}{2} \frac{|Q|}{\sqrt{4 \pi G}} \frac{p^2}{Q^2}<0.
\end{align}
Interestingly, the second order corrections in Eqs.~\eqref{lower_b}-\eqref{extremality} precisely match with each other, so that the two bounds cannot be satisfied simultaneously
. In other words, the window~\eqref{window} for destroying the black hole may be closed by carefully taking into account the second order correction to the total energy. Again, we emphasize that the bound~\eqref{lower_b} does not apply directly to the extremal black hole $(\varepsilon=0)$ and so it is not clear yet if the cosmic censorship conjecture is satisfied for the extremal case, but our observation motivates us to generalize the analysis in Ref.~\cite{Sorce:2017dst} to the extremal black hole.
We leave this interesting problem to the future work.
 
\begin{acknowledgments}
D.Y. would like to thank Hirotaka Yoshino for useful discussion in the 24th workshop on spacetime singularities in Japan. 
K. I. and D. Y. are supported by Grant-Aid for Scientific Research from Ministry of Education,
Science, Sports and Culture of Japan (No. JP21H05189). K. I. and T. N. are supported by JSPS KAKENHI Grant No. JP20H01902.
All of the authors are supported by JSPS Bilateral Joint
Research Projects (JSPS-DST collaboration) (JPJSBP120227705). 
K. I. is also supported by JSPS(JP21H05182).  
T. N. is also supported by JSPS KAKENHI Grant No. JP22H01220 and MEXT KAKENHI Grant No. JP21H05184 and No. JP23H04007.
D.Y. is supported by JSPS KAKENHI Grant No.~JP20K14469.
\end{acknowledgments}

\appendix
\section{Detailed Expression for the Higher Derivative Corrections}
\label{sec.A}

\subsection{How to rewrite the curvature square terms}
\label{sec.A1}
We investigate a theory with the higher derivative interactions \eqref{actionhd}, 
where the curvature square terms do not show up in the action.
This is justified, as we have explained below eq.\eqref{actionhd}, 
by the fact that all curvature square terms are rewritten by the other terms. 
In this appendix, we explicitly show how to rewrite the curvature square terms with 
the other higher derivative terms.

The quadratic curvature terms 
\begin{align}
R^2, \quad R_{\mu\nu} R^{\mu\nu}, \quad R_{\mu\nu\rho\sigma}R^{\mu\nu\rho\sigma}
\label{quaRsub}
\end{align}
are expressed by linear combinations of
\begin{align}
R^2, \quad G_{\mu\nu} G^{\mu\nu}, \quad 
GB:= R^2 - 4 R_{\mu\nu} R^{\mu\nu} + R_{\mu\nu\rho\sigma}R^{\mu\nu\rho\sigma}.
\label{quaR}
\end{align}
We investigate quadratic curvature terms \eqref{quaR}, instead of \eqref{quaRsub}.
Suppose that an action includes these quadratic curvature terms \eqref{quaR}
\begin{align}
 \lambda_{1} R^2 + \lambda_{2} G_{\mu\nu} G^{\mu\nu} + \lambda_{3} GB. \label{Rsquare} 
\end{align}
This term can be expressed as  
\begin{align}
&\lambda_{1} R^2 + \lambda_{2} G_{\mu\nu} G^{\mu\nu} + \lambda_{3} GB \nn
& \quad
= \lambda_{1} R^2 + \lambda_{2} \left(G_{\mu\nu} - 8\pi G T_{\mu\nu} \right)\left( G^{\mu\nu} - 8\pi G T^{\mu\nu}\right) \nn
& \qquad
+ \lambda_{3} GB
+ 16 \pi G \lambda_{2} G_{\mu\nu} T^{\mu\nu} - (8\pi G)^2 \lambda_{2} T_{\mu\nu}T^{\mu\nu},
\label{quaR1}
\end{align}
where $T_{\mu\nu}$ is the energy momentum tensor without the higher derivative 
contributions, defined in eq.\eqref{EMTL}. 
Since the Gauss-Bonnet combination $GB$ is topological invariant, 
it does not affect the dynamics and thus we can just ignore it. 
Meanwhile, because $R$ and $G_{\mu\nu} - 8\pi G T_{\mu\nu}$ vanish at the order in $\epsilon^0$, {\it i.e.}, without the higher derivative 
contributions, 
they are  ${\cal O} (\epsilon^1)$ quantities. 
The variations of squares of them gives linear terms of them, that is, they are ${\cal O} (\epsilon^1)$ quantities. 
Suppose the coefficients of quadratic curvature terms $\lambda_i$ is ${\cal O} (\epsilon)$, and then 
the first two terms in the right hand side of eq.\eqref{quaR1} gives ${\cal O} (\epsilon^2)$ contributions in equations of motion, 
which we ignore in our analysis. 
As a result, the quadratic curvature \eqref{Rsquare} is written in the last two terms in the right hand side of eq.\eqref{quaR1}.
The explicit form of each term is 
\begin{align}
&16 \pi G \lambda_{2} G_{\mu\nu} T^{\mu\nu} \nn
&\quad
= 4 \pi G\lambda_{2}\bigl(
- R F_{\mu\nu} F^{\mu\nu} + 4  R_{\mu\nu} F^{\mu \rho} F^{\nu}{}_{\rho} \nn
&\qquad\qquad\qquad
- R H_{\mu\nu} H^{\mu\nu} + 4 R_{\mu\nu} H^{\mu \rho} H^{\nu}{}_{\rho} 
\bigr),
\end{align}
\begin{align}
&- (8\pi G)^2 \lambda_{2} T_{\mu\nu}T^{\mu\nu} \nn
&\quad
=
16 \pi^2 G^2 \lambda_{2} \bigl( F_{\mu\nu} F^{\mu\nu} F_{\rho\sigma} F^{\rho\sigma} - 4 F_{\mu}{}^{\nu} F_{\nu}{}^{\rho} F_{\rho}{}^{\sigma} F_{\sigma}{}^{\mu}
\nn
&\qquad\quad
+ 2 F_{\mu\nu} F^{\mu\nu} H_{\rho\sigma} H^{\rho\sigma} - 8 F_{\mu}{}^{\nu} F_{\nu}{}^{\rho} H_{\rho}{}^{\sigma} H_{\sigma}{}^{\mu}
\nn
&\qquad\quad
+H_{\mu\nu} H^{\mu\nu} H_{\rho\sigma} H^{\rho\sigma}
- 4 H_{\mu}{}^{\nu} H_{\nu}{}^{\rho} H_{\rho}{}^{\sigma} H_{\sigma}{}^{\mu}
\bigr).
\end{align}
More concretely, if one starts with the action with \eqref{Rsquare} terms, 
such effect can be included by replacing the parameters in Eq.~\eqref{actionhd} as follows,
\begin{align}
\alpha_{1} &\rightarrow \alpha_{1} + 16 \pi^2 G^2 \lambda_{2}, \\
\alpha_{2} &\rightarrow \alpha_{2} - 64 \pi^2 G^2\lambda_{2}, \\
\alpha_{3} &\rightarrow \alpha_{3} - 4 \pi G \lambda_{2}, \\
\alpha_{4} &\rightarrow \alpha_{4} + 16 \pi G \lambda_{2}, \\
\gamma_{11} &\rightarrow \gamma_{11} + 32 \pi^2 G^2 \lambda_{2}, \\
\gamma_{21} &\rightarrow \gamma_{21} - 128 \pi^2 G^2 \lambda_{2}, \\
\gamma_{3} &\rightarrow \gamma_{3} - 4 \pi G \lambda_{2}, \\
\gamma_{4} &\rightarrow \gamma_{4} + 16 \pi G \lambda_{2}, \\
\zeta_{1} &\rightarrow \zeta_{1} + 16 \pi^2 G^2 \lambda_{2},  \\
\zeta_{2} &\rightarrow \zeta_{2} - 64 \pi^2 G^2 \lambda_{2}.
\end{align}

\subsection{Source Terms in the Equations of Motion}
\label{sec.A3}
In this section, we represent the detailed expressions for $\delta T_{\mu\nu}, \delta J^{\mu}_{F},$ and $\delta J^{\mu}_{H}$ appearing in the right hand side of Eqs.~\eqref{eomshd1} to \eqref{eomshd3}. The expressions are as follows:
\begin{align}
&\delta T_{\mu\nu}=
\alpha_{1} {\cal T}_{1,\mu\nu}[F,F,F,F] + \alpha_{2} {\cal T}_{2,\mu\nu}[F,F,F,F] \nn
&\quad
+\beta_{1} {\cal T}_{1,\mu\nu}[F,F,F,H] + \beta_{2} {\cal T}_{2,\mu\nu}[F,F,F,H] \nn
&\quad
+\gamma_{11} {\cal T}_{1,\mu\nu}[F,F,H,H] + \gamma_{12} {\cal T}_{1,\mu\nu}[F,H,F,H]\nn
&\quad 
+\gamma_{21} {\cal T}_{2,\mu\nu}[F,F,H,H] + \gamma_{22} {\cal T}_{2,\mu\nu}[F,H,F,H]\nn
&\quad
+\kappa_{1} {\cal T}_{1,\mu\nu}[F,H,H,H] + \kappa_{2} {\cal T}_{2,\mu\nu}[F,H,H,H] \nn
&\quad
+\zeta_{1} {\cal T}_{1,\mu\nu}[H,H,H,H] + \zeta_{2} {\cal T}_{2,\mu\nu}[H,H,H,H] \nn
&\quad
+ \alpha_{3} {\cal T}_{3,\mu\nu}[F,F] + \alpha_{4} {\cal T}_{4,\mu\nu}[F,F] + \alpha_{5} {\cal T}_{5,\mu\nu}[F,F]\nn
&\quad
+ \beta_{3} {\cal T}_{3,\mu\nu}[F,H] + \beta_{4} {\cal T}_{4,\mu\nu}[F,H] + \beta_{5} {\cal T}_{5,\mu\nu}[F,H]\nn
&\quad
 + \gamma_{3} {\cal T}_{3,\mu\nu}[H,H] + \gamma_{4} {\cal T}_{4,\mu\nu}[H,H] + \gamma_{5} {\cal T}_{5,\mu\nu}[H,H],
\end{align}
\begin{align}
&\delta J_{F}^{\mu}
 = 
 \nabla_{\nu} \Bigl( 
8\alpha_1 {\cal J}_1^{\mu\nu}[F,F,F]+8\alpha_2 {\cal J}_2^{\mu\nu}[F,F,F]\nn
&\qquad\qquad
+\beta_1\bigl( 2 {\cal J}_1^{\mu\nu}[F,F,H] +4 {\cal J}_1^{\mu\nu}[F,H,F]\bigr)\nn
&\qquad\qquad
+\beta_2\bigl( 4 {\cal J}_2^{\mu\nu}[F,F,H] +2 {\cal J}_2^{\mu\nu}[F,H,F]\bigr)\nn
&\quad \qquad
+4\gamma_{11} {\cal J}_1^{\mu\nu}[H,H,F] 
+4\gamma_{12} {\cal J}_1^{\mu\nu}[F,H,H] \nn
&\quad \qquad
+4\gamma_{21} {\cal J}_2^{\mu\nu}[F,H,H] 
+4\gamma_{22} {\cal J}_2^{\mu\nu}[H,F,H] \nn
&\quad \qquad
+2 \kappa_{1} {\cal J}_1^{\mu\nu}[H,H,H] 
+2 \kappa_{2} {\cal J}_2^{\mu\nu}[H,H,H] \nn
&\quad
+4  \alpha_{3} R F^{\nu\mu}
+ 4 \alpha_{4} R^{[\nu|}{}_{\rho} F^{\rho|\mu]}
+ 4 \alpha_{5} R^{\nu\mu\rho\sigma} F_{\rho\sigma}\nn
&\quad
+2 \beta_{3} R H^{\nu\mu}
+2 \beta_{4} R^{[\nu|}{}_{\rho} H^{\rho|\nu]}
+2 \beta_{5} R^{\nu\mu\rho\sigma} H_{\rho\sigma}
\Bigr),
\end{align}
\begin{align}
&\delta J_{H}^{\mu}
 = 
 \nabla_{\nu} \Bigl( 
8\zeta_1 {\cal J}_1^{\mu\nu}[H,H,H]+8\zeta_2 {\cal J}_2^{\mu\nu}[H,H,H]\nn
&\qquad\qquad
+\kappa_1\bigl( 2 {\cal J}_1^{\mu\nu}[H,H,F] +4 {\cal J}_1^{\mu\nu}[H,F,H]\bigr)\nn
&\qquad\qquad
+\kappa_2\bigl( 4 {\cal J}_2^{\mu\nu}[H,H,F] +2 {\cal J}_2^{\mu\nu}[H,F,H]\bigr)\nn
&\quad \qquad
+4\gamma_{11} {\cal J}_1^{\mu\nu}[F,F,H] 
+4\gamma_{12} {\cal J}_1^{\mu\nu}[H,F,F] \nn
&\quad \qquad
+4\gamma_{21} {\cal J}_2^{\mu\nu}[H,F,F] 
+4\gamma_{22} {\cal J}_2^{\mu\nu}[F,H,F] \nn
&\quad \qquad
+2 \beta_{1} {\cal J}_1^{\mu\nu}[F,F,F] 
+2 \beta_{2} {\cal J}_2^{\mu\nu}[F,F,F] \nn
&\quad
+4  \gamma_{3} R H^{\nu\mu}
+ 4 \gamma_{4} R^{[\nu|}{}_{\rho} G^{\rho|\mu]}
+ 4 \gamma_{5} R^{\nu\mu\rho\sigma} G_{\rho\sigma}\nn
&\quad
+2 \beta_{3} R F^{\nu\mu}
+2 \beta_{4} R^{[\nu|}{}_{\rho} F^{\rho|\nu]}
+2 \beta_{5} R^{\nu\mu\rho\sigma} F_{\rho\sigma}
\Bigr),
\end{align}
where ${\cal T}_{i,\mu\nu}[A,B,C,D]$, ${\cal T}_{i,\mu\nu}[A,B]$ and ${\cal J}_i^{\mu\nu}[A,B,C]$ are functionals of antisymmetric tensors $A_{\mu\nu}$, $B_{\mu\nu}$, $C_{\mu\nu}$, $D_{\mu\nu}$ defined as 
\begin{align}
&{\cal T}_{1,\mu\nu}[A,B,C,D] = 
  g_{\mu\nu} A_{\alpha\beta} B^{\beta\alpha} C_{\gamma\lambda}D^{\lambda\gamma}\nn
&\qquad
-4 A_{(\mu|\alpha}B^{\alpha}{}_{|\nu)} C_{\beta\gamma}D^{\gamma\beta}-4 C_{(\mu|\alpha}D^{\alpha}{}_{|\nu)} A_{\beta\gamma}B^{\gamma\beta},
\\
&{\cal T}_{2,\mu\nu}[A,B,C,D]= g_{\mu\nu} A_{\alpha\beta}B^{\beta\gamma}C_{\gamma\lambda}D^{\lambda\alpha}\nn
&\qquad
- 2 A_{(\mu|\alpha}B^{\alpha\beta}C_{\beta\gamma}D^\gamma{}_{\nu)}
- 2 B_{(\mu|\alpha}C^{\alpha\beta}D_{\beta\gamma}A^\gamma{}_{\nu)}\nn
&\qquad
- 2 C_{(\mu|\alpha}D^{\alpha\beta}A_{\beta\gamma}B^\gamma{}_{\nu)}
- 2 D_{(\mu|\alpha}A^{\alpha\beta}B_{\beta\gamma}C^\gamma{}_{\nu)},
\\
&{\cal T}_{3,\mu\nu}[A,B]  \nn
&\quad
=
-R A_{\alpha\beta}B^{\beta\alpha} g_{\mu\nu}
+ 2 A_{\alpha\beta}B^{\beta\alpha} R_{\mu\nu} 
+ 4 R A_{(\mu|\beta}B^{\beta}{}_{|\nu)}\nn
&\qquad
- 2 \nabla_{\mu} \nabla_{\nu} ( A_{\alpha\beta}B^{\beta\alpha})
+ 2 g_{\mu\nu} \Box (A_{\alpha\beta}B^{\beta\alpha}),
\\
&{\cal T}_{4,\mu\nu}[A,B] 
= - R_{\rho\sigma} A^{\rho\alpha}B_\alpha{}^{\sigma} g_{\mu\nu}
+ 2 R_{(\mu}{}^{\rho}  A_{\nu)\alpha}B^\alpha{}_{\rho}\nn
&\qquad
+ 2 R_{(\mu}{}^{\rho}  B_{\nu)\alpha}A^\alpha{}_{\rho}
- 2 R_{\rho\sigma} A^{\rho}{}_{(\mu} B^{\sigma}{}_{\nu)}
\nn
&\qquad
+ \Box (A_{(\mu|\alpha} B^\alpha{}_{|\nu)})
+ \nabla_{\rho} \nabla_{\sigma}( A^{\rho\alpha}B_\alpha{}^{\sigma}) g_{\mu\nu}
\nn
&\qquad
- \nabla_{\rho} \nabla_{(\mu} \left( A_{\nu)\alpha}B^{\alpha\rho} \right)
- \nabla_{\rho} \nabla_{(\mu} \left( B_{\nu)\alpha}A^{\alpha\rho} \right),
\\
&{\cal T}_{5,\mu\nu}[A,B] 
=
R_{\rho\sigma\lambda\tau} A^{\rho\sigma} B^{\lambda\tau} g_{\mu\nu}\nn
&\qquad
- 3 R_{(\mu}{}^{\rho\sigma\tau} A_{\nu) \rho} B_{\sigma\tau} 
- 3 R_{(\mu}{}^{\rho\sigma\tau} B_{\nu) \rho} A_{\sigma\tau} \nn
&\qquad
- 2 \nabla_{\rho} \nabla_{\sigma} \left( A_{(\mu}{}^{\rho} B_{\nu)}{}^{\sigma} + B_{(\mu}{}^{\rho} A_{\nu)}{}^{\sigma}\right) , \\
&{\cal J}_1^{\mu\nu}[A,B,C]=A_{\alpha \beta}B^{\beta\alpha}C^{\mu\nu}, \\
&{\cal J}_2^{\mu\nu}[A,B,C]=A^{[\mu|}{}_\alpha B^{\alpha\beta}C_\beta{}^{|\nu]}.
\end{align}

\begin{widetext}

\section{Field Redefinition}
\label{sec.B}

Field redefinition changes the description of the theory, for instance, changes the coupling constants, 
but the physical phenomenon remains the same. 
Although the theory we consider can be rewritten in a different description by a field redefinition, 
in this appendix we explicitly confirm that the obtained results are the same. 
This is obvious, but it becomes a cross-check of the validity of our analysis, and 
may bring a new perspective on the analysis of theories with the higher derivative terms.

Let us investigate the perturbative redefinition of the metric field $g_{\mu\nu}$ to $\hat{g}_{\mu\nu}$ through, 
\begin{align}
 g_{\mu\nu} = \hat{g}_{\mu\nu} +  \delta g_{\mu\nu},\label{fieldredefinition}
\end{align}
with $\delta g_{\mu\nu}$ defined by
\begin{align}
 \delta g_{\mu\nu} =& \epsilon_{1} R_{\mu\nu} + \epsilon_{2} g_{\mu\nu} R + 8 \pi G \epsilon_{3} F_{\mu \rho} F_{\nu}{}^{\rho}
+ 8 \pi G \epsilon_{4} g_{\mu\nu} F_{\rho\sigma} F^{\rho\sigma} 
 +8 \pi G \epsilon_{5} F_{\mu \rho} P_{\nu}{}^{\rho} \nn
&
+ 8 \pi G \epsilon_{6} g_{\mu\nu} F_{\rho\sigma} P^{\rho\sigma}
+8 \pi G \epsilon_{7} P_{\mu \rho} P_{\nu}{}^{\rho}
+ 8 \pi G \epsilon_{8} g_{\mu\nu} P_{\rho\sigma} P^{\rho\sigma}
\label{defFRD}
,
\end{align}
where the coefficients are regarded as ${\cal O}(\epsilon)$. 
One can replace the metric $g_{\mu\nu}$ in Eq.~\eqref{defFRD} with $\hat{g}_{\mu\nu}$ because the difference is higher order in $\epsilon$.
Let us investigate the influence of the field redefinition~\eqref{fieldredefinition} in the black hole solutions and the falling conditions.

\subsection{Black hole solutions}
\label{AppA1a}

Through this transformation, the action~\eqref{actionhd},
expressing the dependence of the metric and the parameters as $S[g_{\mu\nu}, \alpha_{i}]$, is modified to 
\begin{align}
 S[g_{\mu\nu}, \alpha_{i}] &= S[\hat{g}_{\mu\nu}, \alpha_{i}] - \int d^4x \frac{\sqrt{- g}}{16 \pi G} \left( R^{\mu\nu} - \frac{1}{2} g^{\mu\nu} R - 8 \pi G T^{\mu\nu} \right) \delta g_{\mu\nu} + {\cal O}(\epsilon^2) \nonumber \\
&= S[\hat{g}_{\mu\nu}, \hat{\alpha}_{i}] + \int d^4 x  \frac{\sqrt{-g}}{16 \pi G} \left[ 
 - \epsilon_{1} \left( R_{\mu\nu} - \frac{1}{2} R g_{\mu\nu} - 8 \pi G T_{\mu\nu} \right)^2 
+ \left( \frac{\epsilon_{1}}{2} + \epsilon_{2} \right) R^2 \right]
+ {\cal O}(\epsilon^2). \label{actionghat}
\end{align}
\end{widetext}
Here the parameters $\hat{\alpha}_{i}$ are defined as follows:
\begin{align}
 \hat{\alpha}_{1} & = \alpha_{1} - \pi G (\epsilon_{1} + \epsilon_{3}), \label{a1e}\\ 
 \hat{\alpha}_{2} & = \alpha_{2} + 4 \pi G (\epsilon_{1} + \epsilon_{3}), \\
 \hat{\alpha}_{3} & = \alpha_{3} + \frac{1}{8}(\epsilon_{1} + 2 \epsilon_{3} + 4 \epsilon_{4}),\\
 \hat{\alpha}_{4} & = \alpha_{4} - \frac{1}{2}( \epsilon_{1} + \epsilon_{3}),\\
 \hat{\alpha}_{5} & = \alpha_{5},\\
 \hat{\beta}_{1} & = \beta_{1} - \pi G \epsilon_{5}, \\
 \hat{\beta}_{2} & = \beta_{2} + 4 \pi G \epsilon_{5},\\
 \hat{\beta}_{3} & = \beta_{3} + \frac{1}{4} (\epsilon_{5} + 2 \epsilon_{6}), \\
 \hat{\beta}_{4} & = \beta_{4} - \frac{1}{2} \epsilon_{5},\\
 \hat{\beta}_{5} & = \beta_{5},\\
\hat{\gamma}_{11} &= \gamma_{11} - \pi G (2 \epsilon_{1} + \epsilon_{3}+\epsilon_{7}),\\
\hat{\gamma}_{12} &= \gamma_{12}, \\
\hat{\gamma}_{21} &= \gamma_{21} + 4\pi G (2 \epsilon_{1} + \epsilon_{3} + \epsilon_{7}),\\
\hat{\gamma}_{22} &= \gamma_{22}, 
\end{align}
\begin{align}
\hat{\gamma}_{3} &= \gamma_{3} + \frac{1}{8}( \epsilon_{1} + 2\epsilon_{7} + 4 \epsilon_{8}),\\
\hat{\gamma}_{4} &= \gamma_{4} - \frac{1}{2}( \epsilon_{1} + \epsilon_{7}),\\ 
\hat{\gamma}_{5} &= \gamma_{5},\\ 
\hat{\kappa}_{1} &= \kappa_{1} - \pi G \epsilon_{5}, \\
\hat{\kappa}_{2} &= \kappa_{2} + 4 \pi G \epsilon_{5}, \\
\hat{\zeta}_{1} &= \zeta_{1} - \pi G (\epsilon_{1} + \epsilon_{7}), \\
\hat{\zeta}_{2} &= \zeta_{2} + 4 \pi G (\epsilon_{1} + \epsilon_{7}). \label{z2e}
\end{align}
The second term in Eq.~\eqref{actionghat} is irrelevant for our analysis because it consists of squares of the background equations of motion. Thus, the relevant part of the action for $\hat{g}_{\mu\nu}$ can be obtained by replacing the parameters in the original action \eqref{actionhd} with the hatted ones defined above.

The transformations for the coupling constants are described by the six independent combinations of the parameters,
\begin{align}
\epsilon_{1} + \epsilon_{3}, \, \epsilon_{3} + 4 \epsilon_{4}, \, \epsilon_{5}, \, \epsilon_{6}, \, \epsilon_{1} + \epsilon_{7}, \, \epsilon_{7} + 4 \epsilon_{8}.
\label{FRDep}
\end{align}
One can see that the following combinations are invariant under the field redefinition~\eqref{fieldredefinition},
\begin{align}
& \alpha_{1} - 2 \pi G \alpha_{4}, \\
&\alpha_{2} + 8 \pi G \alpha_{4},\\
&\beta_{1} - 2 \pi G \beta_{4},\\
&\beta_{2} + 8 \pi G \beta_{4},\\
&\gamma_{11} - 2 \pi G (\alpha_{4} + \gamma_{4}),\\
&\gamma_{21} + 8 \pi G (\alpha_{4} + \gamma_{4}),\\
&\kappa_{1} - 2 \pi G \beta_{4},\\
&\kappa_{2} + 8 \pi G \beta_{4},\\
&\zeta_{1} - 2 \pi G \gamma_{4},\\
&\zeta_{2} + 8 \pi G \gamma_{4},
\end{align}
as well as $\alpha_{5}, \beta_{5}, \gamma_{12}, \gamma_{22}$ and $\gamma_{5}$. The parameters $\alpha, \beta, \gamma, \kappa, \zeta$, appearing in the extremal condition~\eqref{generalext}, are invariant thus under the field redefinition. 

Next let us investigate the position of the horizon. 
The horizon is characterized by $f(r)=0$.
\footnote{
In theories with non-canonical kinetic terms, superluminal modes can exist. 
Thus, an event horizon may not be a null surface. 
However, in static or stationary case, any horizon is expected to be a Killing horizon~\cite{Izumi:2014loa,Reall:2014pwa,Reall:2021voz}. 
} 
Through the field redefinition, $\hat{f} := - \hat{g}_{tt}$ can be calculated as
\begin{align}
 \hat{f} = f - \delta f
\end{align}
with
\begin{align} 
\delta f &= - \delta g_{tt} \nn
& =- f(r) \frac{G}{4 \pi r^4} \bigl(
Q^2 (\epsilon_{1} + 2 \epsilon_{3} + 4 \epsilon_{4}) + 2 Q P (\epsilon_{5} +2  \epsilon_{6})\nn
&\hspace{30mm}
+ P^2 (\epsilon_{1} + 2 \epsilon_{7} + 4 \epsilon_{8} )
\bigr).
\end{align}
Although $\hat{f}(r)$ does not coincide with $f(r)$ itself, 
a solution of $f(r)=0$ is the solution of $\hat{f}(r)=0$, that is, 
the position of the horizon is invariant. 

We note that it does not imply the invariance of the areal radius of the horizon. 
As we can see from eq.\eqref{correctedrH}, $r_{\text{H}}$ is not invariant under the field redefinition.
The reason is as follows. 
After the field redefinition, the function forms of metric change.
Due to the spherical symmetry, metric after the field redefinition is written as 
\begin{align}
\hat g_{\mu\nu} dx^{\mu} dx^{\nu} = - \hat f(r) dt^2 + \frac{\hat h(r)}{\hat f(r)}dr^2 + \hat r(r)^2 d\Omega^2.\label{ansatzbarg}
\end{align} 
The invariance of the horizon position means that, in this coordinate with $r$, the locus of horizon is the same. However, $r$ does not coincide with the areal radius $\hat r(r)$ in metric~\eqref{ansatzbarg}. 
The areal radius is $\hat r (r)$, which is calculated as 
\begin{align}
\hat r^2 = r^2 - \epsilon_1 \frac{G(Q^2+P^2)}{4\pi r^2} + \epsilon_4 \frac{G Q^2}{\pi r^2} + \epsilon_6 \frac{G QP}{\pi r^2} + \epsilon_8 \frac{G P^2}{\pi r^2}.
\end{align} 
This gives us 
\begin{align}
&\hat{r} = r - \epsilon_1 \frac{G(Q^2+P^2)}{8\pi r^3}+\epsilon_4 \frac{G Q^2}{2\pi r^3}+\epsilon_6 \frac{G QP}{2\pi r^3} \nn
&\hspace{30mm}
+\epsilon_8 \frac{G P^2}{2\pi r^3} + {\cal O}\left(\epsilon^2\right).
\end{align}
Thus, $\hat{r}_{\text{H}}$, the areal radius of the horizon for $\hat{g}_{\mu\nu}$, can be obtained as
\begin{align}
 &\hat{r}_{\mathrm{H}} = r_{\mathrm{H}} - \epsilon_1 \frac{1}{2 \bar{r}_{\mathrm{H}}}+\epsilon_4 \frac{G Q^2}{2\pi \bar{r}_{\mathrm{H}}^3}+\epsilon_6 \frac{G QP}{2\pi r_{\mathrm{H}}^3}
\nn
&\hspace{30mm}
+\epsilon_8 \frac{G P^2}{2\pi \bar{r}_{\mathrm{H}}^3} + {\cal O}\left(\epsilon^2\right),
\end{align}
where we have used
\begin{align}
r_{\text{H}} =  \bar{r}_{\text{H}} + {\cal O}(\epsilon).
\end{align}
This shift of the areal radius of the horizon is consistent with
Eq.~\eqref{correctedrH}.

\subsection{Falling conditions}
\label{AppFC}

The action~\eqref{Actionparticle} gives a theory for a charged particle where 
the particle is minimally coupled with gravity. 
One may consider a case where the charged particle has the higher order derivative coupling
\begin{align}
 S^{\text{test}}[x^{\mu}] = \int d\tau \biggl[& - m \sqrt{- \left(g_{\mu\nu}(x)+ \delta g_{\mu\nu}^{(p)}(x) \right)  \dot{x}^{\mu} \dot{x}^{\nu}} \notag\\
& + q A_{\mu}(x) \dot{x}^{\mu}
+ p B_{\mu}(x) \dot{x}^{\mu}
 \biggr],
 \label{Actionpnm}
\end{align}
where
\begin{align}
 \delta g^{(p)}_{\mu\nu} =&  \mu_{1} R_{\mu\nu} + \mu_{2} g_{\mu\nu} R + 8 \pi G \mu_{3} F_{\mu \rho} F^{\nu \rho} \nn
 &
 + 8 \pi G \mu_{4} g_{\mu\nu} F_{\rho\sigma} F^{\rho\sigma} 
 +8 \pi G \mu_{5} F_{\mu \rho} P^{\nu \rho} \nn
 &+ 8 \pi G \mu_{6} g_{\mu\nu} F_{\rho\sigma} P^{\rho\sigma} 
+8 \pi G \mu_{7} P_{\mu \rho} P^{\nu \rho} \nn
 &+ 8 \pi G \mu_{8} g_{\mu\nu} P_{\rho\sigma} P^{\rho\sigma}.
\end{align}
Here, $\mu_i$'s are ${\cal O}(\epsilon)$ constants. 
Although this theory is reduced to that with the action~\eqref{Actionparticle} by the field redefinition~\eqref{FRDep} with $\epsilon_i =-\mu_i$, 
we can directly show that falling conditions becomes the same as those obtained in Sec.~\ref{sec.2}. 

The analysis is very simple. 
We have a degree of freedom for the parameter $\tau$ of the worldline $x(\tau)$. 
Instead of $g_{\mu\nu} \dot x^\mu \dot x^\nu=-1$, 
we set the parameter $\tau$ by $(g_{\mu\nu}+\delta g^{(p)}_{\mu\nu}) \dot x^\mu \dot x^\nu=-1$. 
Then, all the equations become  those in Sec.~\ref{sec.2} with replacing $g_{\mu\nu}$ by $g_{\mu\nu}+\delta g^{(p)}_{\mu\nu}$. 
Though we used the areal radius in Sec.~\ref{sec.2}, the same result holds with any radial coordinate. Thus, we can obtain the result simply by replacing $f$ and $h$ with the corresponding components of $g_{\mu\nu}+\delta g^{(p)}_{\mu\nu}$. As similar to the discussion in the previous subsection, the event horizon for $g_{\mu\nu}+\delta g^{(p)}_{\mu\nu}$ is still located at $r = r_{\text{H}}$ and hence the value of the potential hight at the event horizon, the right hand side of Eq.~\eqref{falloff}, is unchanged. In addition, since $\delta g^{(p)}_{\mu\nu}$ disappears at $r \rightarrow \infty$, the definition of $E$ is also unchanged.
Therefore, the falling conditions with action~\eqref{Actionpnm} become the same as those with action~\eqref{Actionparticle} whose metric is $g_{\mu\nu}+\delta g^{(p)}_{\mu\nu}$.

\newpage
\bibliography{ref}

\begin{thebibliography}{26}%
\makeatletter
\providecommand \@ifxundefined [1]{%
 \@ifx{#1\undefined}
}%
\providecommand \@ifnum [1]{%
 \ifnum #1\expandafter \@firstoftwo
 \else \expandafter \@secondoftwo
 \fi
}%
\providecommand \@ifx [1]{%
 \ifx #1\expandafter \@firstoftwo
 \else \expandafter \@secondoftwo
 \fi
}%
\providecommand \natexlab [1]{#1}%
\providecommand \enquote  [1]{``#1''}%
\providecommand \bibnamefont  [1]{#1}%
\providecommand \bibfnamefont [1]{#1}%
\providecommand \citenamefont [1]{#1}%
\providecommand \href@noop [0]{\@secondoftwo}%
\providecommand \href [0]{\begingroup \@sanitize@url \@href}%
\providecommand \@href[1]{\@@startlink{#1}\@@href}%
\providecommand \@@href[1]{\endgroup#1\@@endlink}%
\providecommand \@sanitize@url [0]{\catcode `\\12\catcode `\$12\catcode
  `\&12\catcode `\#12\catcode `\^12\catcode `\_12\catcode `\%12\relax}%
\providecommand \@@startlink[1]{}%
\providecommand \@@endlink[0]{}%
\providecommand \url  [0]{\begingroup\@sanitize@url \@url }%
\providecommand \@url [1]{\endgroup\@href {#1}{\urlprefix }}%
\providecommand \urlprefix  [0]{URL }%
\providecommand \Eprint [0]{\href }%
\providecommand \doibase [0]{https://doi.org/}%
\providecommand \selectlanguage [0]{\@gobble}%
\providecommand \bibinfo  [0]{\@secondoftwo}%
\providecommand \bibfield  [0]{\@secondoftwo}%
\providecommand \translation [1]{[#1]}%
\providecommand \BibitemOpen [0]{}%
\providecommand \bibitemStop [0]{}%
\providecommand \bibitemNoStop [0]{.\EOS\space}%
\providecommand \EOS [0]{\spacefactor3000\relax}%
\providecommand \BibitemShut  [1]{\csname bibitem#1\endcsname}%
\let\auto@bib@innerbib\@empty
\bibitem [{\citenamefont {Penrose}(1969)}]{Penrose:1969pc}%
  \BibitemOpen
  \bibfield  {author} {\bibinfo {author} {\bibfnamefont {R.}~\bibnamefont
  {Penrose}},\ }\bibfield  {title} {\bibinfo {title} {{Gravitational collapse:
  The role of general relativity}},\ }\href
  {https://doi.org/10.1023/A:1016578408204} {\bibfield  {journal} {\bibinfo
  {journal} {Riv. Nuovo Cim.}\ }\textbf {\bibinfo {volume} {1}},\ \bibinfo
  {pages} {252} (\bibinfo {year} {1969})}\BibitemShut {NoStop}%
\bibitem [{\citenamefont {Wald}(1974)}]{Wald:1974hkz}%
  \BibitemOpen
  \bibfield  {author} {\bibinfo {author} {\bibfnamefont {R.}~\bibnamefont
  {Wald}},\ }\bibfield  {title} {\bibinfo {title} {{Gedanken experiments to
  destroy a black hole}},\ }\href
  {https://doi.org/10.1016/0003-4916(74)90125-0} {\bibfield  {journal}
  {\bibinfo  {journal} {Annals Phys.}\ }\textbf {\bibinfo {volume} {82}},\
  \bibinfo {pages} {548} (\bibinfo {year} {1974})}\BibitemShut {NoStop}%
\bibitem [{\citenamefont {Cohen}\ and\ \citenamefont
  {Gautreau}(1979)}]{Cohen:1979zzb}%
  \BibitemOpen
  \bibfield  {author} {\bibinfo {author} {\bibfnamefont {J.~M.}\ \bibnamefont
  {Cohen}}\ and\ \bibinfo {author} {\bibfnamefont {R.}~\bibnamefont
  {Gautreau}},\ }\bibfield  {title} {\bibinfo {title} {{Naked singularities,
  event horizons, and charged particles}},\ }\href
  {https://doi.org/10.1103/PhysRevD.19.2273} {\bibfield  {journal} {\bibinfo
  {journal} {Phys. Rev. D}\ }\textbf {\bibinfo {volume} {19}},\ \bibinfo
  {pages} {2273} (\bibinfo {year} {1979})}\BibitemShut {NoStop}%
\bibitem [{\citenamefont {Needham}(1980)}]{Needham:1980fb}%
  \BibitemOpen
  \bibfield  {author} {\bibinfo {author} {\bibfnamefont {T.}~\bibnamefont
  {Needham}},\ }\bibfield  {title} {\bibinfo {title} {{COSMIC CENSORSHIP AND
  TEST PARTICLES}},\ }\href {https://doi.org/10.1103/PhysRevD.22.791}
  {\bibfield  {journal} {\bibinfo  {journal} {Phys. Rev. D}\ }\textbf {\bibinfo
  {volume} {22}},\ \bibinfo {pages} {791} (\bibinfo {year} {1980})}\BibitemShut
  {NoStop}%
\bibitem [{\citenamefont {Semiz}(1990)}]{Semiz:1990fm}%
  \BibitemOpen
  \bibfield  {author} {\bibinfo {author} {\bibfnamefont {I.}~\bibnamefont
  {Semiz}},\ }\bibfield  {title} {\bibinfo {title} {{Dyon black holes do not
  violate cosmic censorship}},\ }\href
  {https://doi.org/10.1088/0264-9381/7/3/009} {\bibfield  {journal} {\bibinfo
  {journal} {Class. Quant. Grav.}\ }\textbf {\bibinfo {volume} {7}},\ \bibinfo
  {pages} {353} (\bibinfo {year} {1990})}\BibitemShut {NoStop}%
\bibitem [{\citenamefont {Bekenstein}\ and\ \citenamefont
  {Rosenzweig}(1994)}]{Bekenstein:1994nx}%
  \BibitemOpen
  \bibfield  {author} {\bibinfo {author} {\bibfnamefont {J.~D.}\ \bibnamefont
  {Bekenstein}}\ and\ \bibinfo {author} {\bibfnamefont {C.}~\bibnamefont
  {Rosenzweig}},\ }\bibfield  {title} {\bibinfo {title} {{Stability of the
  black hole horizon and the Landau ghost}},\ }\href
  {https://doi.org/10.1103/PhysRevD.50.7239} {\bibfield  {journal} {\bibinfo
  {journal} {Phys. Rev. D}\ }\textbf {\bibinfo {volume} {50}},\ \bibinfo
  {pages} {7239} (\bibinfo {year} {1994})},\ \Eprint
  {https://arxiv.org/abs/gr-qc/9406024} {arXiv:gr-qc/9406024} \BibitemShut
  {NoStop}%
\bibitem [{\citenamefont {Semiz}(2011)}]{Semiz:2005gs}%
  \BibitemOpen
  \bibfield  {author} {\bibinfo {author} {\bibfnamefont {I.}~\bibnamefont
  {Semiz}},\ }\bibfield  {title} {\bibinfo {title} {{Dyonic Kerr-Newman black
  holes, complex scalar field and cosmic censorship}},\ }\href
  {https://doi.org/10.1007/s10714-010-1108-z} {\bibfield  {journal} {\bibinfo
  {journal} {Gen. Rel. Grav.}\ }\textbf {\bibinfo {volume} {43}},\ \bibinfo
  {pages} {833} (\bibinfo {year} {2011})},\ \Eprint
  {https://arxiv.org/abs/gr-qc/0508011} {arXiv:gr-qc/0508011} \BibitemShut
  {NoStop}%
\bibitem [{\citenamefont {Drummond}\ and\ \citenamefont
  {Hathrell}(1980)}]{Drummond:1979pp}%
  \BibitemOpen
  \bibfield  {author} {\bibinfo {author} {\bibfnamefont {I.~T.}\ \bibnamefont
  {Drummond}}\ and\ \bibinfo {author} {\bibfnamefont {S.~J.}\ \bibnamefont
  {Hathrell}},\ }\bibfield  {title} {\bibinfo {title} {{QED Vacuum Polarization
  in a Background Gravitational Field and Its Effect on the Velocity of
  Photons}},\ }\href {https://doi.org/10.1103/PhysRevD.22.343} {\bibfield
  {journal} {\bibinfo  {journal} {Phys. Rev. D}\ }\textbf {\bibinfo {volume}
  {22}},\ \bibinfo {pages} {343} (\bibinfo {year} {1980})}\BibitemShut
  {NoStop}%
\bibitem [{\citenamefont {Kats}\ \emph {et~al.}(2007)\citenamefont {Kats},
  \citenamefont {Motl},\ and\ \citenamefont {Padi}}]{Kats:2006xp}%
  \BibitemOpen
  \bibfield  {author} {\bibinfo {author} {\bibfnamefont {Y.}~\bibnamefont
  {Kats}}, \bibinfo {author} {\bibfnamefont {L.}~\bibnamefont {Motl}},\ and\
  \bibinfo {author} {\bibfnamefont {M.}~\bibnamefont {Padi}},\ }\bibfield
  {title} {\bibinfo {title} {{Higher-order corrections to mass-charge relation
  of extremal black holes}},\ }\href
  {https://doi.org/10.1088/1126-6708/2007/12/068} {\bibfield  {journal}
  {\bibinfo  {journal} {JHEP}\ }\textbf {\bibinfo {volume} {12}},\ \bibinfo
  {pages} {068}},\ \Eprint {https://arxiv.org/abs/hep-th/0606100}
  {arXiv:hep-th/0606100} \BibitemShut {NoStop}%
\bibitem [{\citenamefont {Cheung}\ \emph {et~al.}(2018)\citenamefont {Cheung},
  \citenamefont {Liu},\ and\ \citenamefont {Remmen}}]{Cheung:2018cwt}%
  \BibitemOpen
  \bibfield  {author} {\bibinfo {author} {\bibfnamefont {C.}~\bibnamefont
  {Cheung}}, \bibinfo {author} {\bibfnamefont {J.}~\bibnamefont {Liu}},\ and\
  \bibinfo {author} {\bibfnamefont {G.~N.}\ \bibnamefont {Remmen}},\ }\bibfield
   {title} {\bibinfo {title} {{Proof of the Weak Gravity Conjecture from Black
  Hole Entropy}},\ }\href {https://doi.org/10.1007/JHEP10(2018)004} {\bibfield
  {journal} {\bibinfo  {journal} {JHEP}\ }\textbf {\bibinfo {volume} {10}},\
  \bibinfo {pages} {004}},\ \Eprint {https://arxiv.org/abs/1801.08546}
  {arXiv:1801.08546 [hep-th]} \BibitemShut {NoStop}%
\bibitem [{\citenamefont {Hamada}\ \emph {et~al.}(2019)\citenamefont {Hamada},
  \citenamefont {Noumi},\ and\ \citenamefont {Shiu}}]{Hamada:2018dde}%
  \BibitemOpen
  \bibfield  {author} {\bibinfo {author} {\bibfnamefont {Y.}~\bibnamefont
  {Hamada}}, \bibinfo {author} {\bibfnamefont {T.}~\bibnamefont {Noumi}},\ and\
  \bibinfo {author} {\bibfnamefont {G.}~\bibnamefont {Shiu}},\ }\bibfield
  {title} {\bibinfo {title} {{Weak Gravity Conjecture from Unitarity and
  Causality}},\ }\href {https://doi.org/10.1103/PhysRevLett.123.051601}
  {\bibfield  {journal} {\bibinfo  {journal} {Phys. Rev. Lett.}\ }\textbf
  {\bibinfo {volume} {123}},\ \bibinfo {pages} {051601} (\bibinfo {year}
  {2019})},\ \Eprint {https://arxiv.org/abs/1810.03637} {arXiv:1810.03637
  [hep-th]} \BibitemShut {NoStop}%
\bibitem [{\citenamefont {Jones}\ and\ \citenamefont
  {McPeak}(2020)}]{Jones:2019nev}%
  \BibitemOpen
  \bibfield  {author} {\bibinfo {author} {\bibfnamefont {C.~R.~T.}\
  \bibnamefont {Jones}}\ and\ \bibinfo {author} {\bibfnamefont
  {B.}~\bibnamefont {McPeak}},\ }\bibfield  {title} {\bibinfo {title} {{The
  Black Hole Weak Gravity Conjecture with Multiple Charges}},\ }\href
  {https://doi.org/10.1007/JHEP06(2020)140} {\bibfield  {journal} {\bibinfo
  {journal} {JHEP}\ }\textbf {\bibinfo {volume} {06}},\ \bibinfo {pages}
  {140}},\ \Eprint {https://arxiv.org/abs/1908.10452} {arXiv:1908.10452
  [hep-th]} \BibitemShut {NoStop}%
\bibitem [{\citenamefont {Leaute}\ and\ \citenamefont
  {Linet}(1982)}]{Leaute:1982sm}%
  \BibitemOpen
  \bibfield  {author} {\bibinfo {author} {\bibfnamefont {B.}~\bibnamefont
  {Leaute}}\ and\ \bibinfo {author} {\bibfnamefont {B.}~\bibnamefont {Linet}},\
  }\bibfield  {title} {\bibinfo {title} {{Selfinteraction of a point charge in
  the Kerr space-time}},\ }\href {https://doi.org/10.1088/0305-4470/15/6/021}
  {\bibfield  {journal} {\bibinfo  {journal} {J. Phys. A}\ }\textbf {\bibinfo
  {volume} {15}},\ \bibinfo {pages} {1821} (\bibinfo {year}
  {1982})}\BibitemShut {NoStop}%
\bibitem [{\citenamefont {Hod}(2002)}]{Hod:2002pm}%
  \BibitemOpen
  \bibfield  {author} {\bibinfo {author} {\bibfnamefont {S.}~\bibnamefont
  {Hod}},\ }\bibfield  {title} {\bibinfo {title} {{Cosmic censorship, area
  theorem, and selfenergy of particles}},\ }\href
  {https://doi.org/10.1103/PhysRevD.66.024016} {\bibfield  {journal} {\bibinfo
  {journal} {Phys. Rev. D}\ }\textbf {\bibinfo {volume} {66}},\ \bibinfo
  {pages} {024016} (\bibinfo {year} {2002})},\ \Eprint
  {https://arxiv.org/abs/gr-qc/0205005} {arXiv:gr-qc/0205005} \BibitemShut
  {NoStop}%
\bibitem [{\citenamefont {Sorce}\ and\ \citenamefont
  {Wald}(2017)}]{Sorce:2017dst}%
  \BibitemOpen
  \bibfield  {author} {\bibinfo {author} {\bibfnamefont {J.}~\bibnamefont
  {Sorce}}\ and\ \bibinfo {author} {\bibfnamefont {R.~M.}\ \bibnamefont
  {Wald}},\ }\bibfield  {title} {\bibinfo {title} {{Gedanken experiments to
  destroy a black hole. II. Kerr-Newman black holes cannot be overcharged or
  overspun}},\ }\href {https://doi.org/10.1103/PhysRevD.96.104014} {\bibfield
  {journal} {\bibinfo  {journal} {Phys. Rev. D}\ }\textbf {\bibinfo {volume}
  {96}},\ \bibinfo {pages} {104014} (\bibinfo {year} {2017})},\ \Eprint
  {https://arxiv.org/abs/1707.05862} {arXiv:1707.05862 [gr-qc]} \BibitemShut
  {NoStop}%
\bibitem [{\citenamefont {Jiang}(2020)}]{Jiang:2019soz}%
  \BibitemOpen
  \bibfield  {author} {\bibinfo {author} {\bibfnamefont {J.}~\bibnamefont
  {Jiang}},\ }\bibfield  {title} {\bibinfo {title} {{Static charged
  Gauss-Bonnet black holes cannot be overcharged by the new version of gedanken
  experiments}},\ }\href {https://doi.org/10.1016/j.physletb.2020.135365}
  {\bibfield  {journal} {\bibinfo  {journal} {Phys. Lett. B}\ }\textbf
  {\bibinfo {volume} {804}},\ \bibinfo {pages} {135365} (\bibinfo {year}
  {2020})},\ \Eprint {https://arxiv.org/abs/1912.10826} {arXiv:1912.10826
  [gr-qc]} \BibitemShut {NoStop}%
\bibitem [{\citenamefont {Wang}\ and\ \citenamefont
  {Jiang}(2020)}]{Wang:2020vpn}%
  \BibitemOpen
  \bibfield  {author} {\bibinfo {author} {\bibfnamefont {X.-Y.}\ \bibnamefont
  {Wang}}\ and\ \bibinfo {author} {\bibfnamefont {J.}~\bibnamefont {Jiang}},\
  }\bibfield  {title} {\bibinfo {title} {{Gedanken experiments at high-order
  approximation: nearly extremal Reissner-Nordstr\"om black holes cannot be
  overcharged}},\ }\href {https://doi.org/10.1007/JHEP05(2020)161} {\bibfield
  {journal} {\bibinfo  {journal} {JHEP}\ }\textbf {\bibinfo {volume} {05}},\
  \bibinfo {pages} {161}},\ \Eprint {https://arxiv.org/abs/2004.12120}
  {arXiv:2004.12120 [hep-th]} \BibitemShut {NoStop}%
\bibitem [{\citenamefont {D\"uzta\c{s}}(2021)}]{Duztas:2021kni}%
  \BibitemOpen
  \bibfield  {author} {\bibinfo {author} {\bibfnamefont {K.}~\bibnamefont
  {D\"uzta\c{s}}},\ }\bibfield  {title} {\bibinfo {title} {{The variational
  method, backreactions, and the absorption probability in Wald type
  problems}},\ }\href {https://doi.org/10.1140/epjc/s10052-021-08879-2}
  {\bibfield  {journal} {\bibinfo  {journal} {Eur. Phys. J. C}\ }\textbf
  {\bibinfo {volume} {81}},\ \bibinfo {pages} {49} (\bibinfo {year} {2021})},\
  \Eprint {https://arxiv.org/abs/2101.07790} {arXiv:2101.07790 [gr-qc]}
  \BibitemShut {NoStop}%
\bibitem [{\citenamefont {Sang}\ and\ \citenamefont
  {Jiang}(2021)}]{Sang:2021xqj}%
  \BibitemOpen
  \bibfield  {author} {\bibinfo {author} {\bibfnamefont {A.}~\bibnamefont
  {Sang}}\ and\ \bibinfo {author} {\bibfnamefont {J.}~\bibnamefont {Jiang}},\
  }\bibfield  {title} {\bibinfo {title} {{Gedanken experiments at high-order
  approximation: Kerr black hole cannot be overspun}},\ }\href
  {https://doi.org/10.1007/JHEP09(2021)095} {\bibfield  {journal} {\bibinfo
  {journal} {JHEP}\ }\textbf {\bibinfo {volume} {09}},\ \bibinfo {pages}
  {095}},\ \Eprint {https://arxiv.org/abs/2108.03454} {arXiv:2108.03454
  [gr-qc]} \BibitemShut {NoStop}%
\bibitem [{\citenamefont {Wang}\ and\ \citenamefont
  {Jiang}(2022)}]{Wang:2022umx}%
  \BibitemOpen
  \bibfield  {author} {\bibinfo {author} {\bibfnamefont {X.-Y.}\ \bibnamefont
  {Wang}}\ and\ \bibinfo {author} {\bibfnamefont {J.}~\bibnamefont {Jiang}},\
  }\bibfield  {title} {\bibinfo {title} {{Gedanken experiments at high-order
  approximations: Kerr-Newman black hole cannot be overcharged and overspun}},\
  }\href {https://doi.org/10.1103/PhysRevD.106.064050} {\bibfield  {journal}
  {\bibinfo  {journal} {Phys. Rev. D}\ }\textbf {\bibinfo {volume} {106}},\
  \bibinfo {pages} {064050} (\bibinfo {year} {2022})}\BibitemShut {NoStop}%
\bibitem [{\citenamefont {Chen}\ \emph {et~al.}(2021)\citenamefont {Chen},
  \citenamefont {Lin}, \citenamefont {Ning},\ and\ \citenamefont
  {Chen}}]{Chen:2020hjm}%
  \BibitemOpen
  \bibfield  {author} {\bibinfo {author} {\bibfnamefont {B.}~\bibnamefont
  {Chen}}, \bibinfo {author} {\bibfnamefont {F.-L.}\ \bibnamefont {Lin}},
  \bibinfo {author} {\bibfnamefont {B.}~\bibnamefont {Ning}},\ and\ \bibinfo
  {author} {\bibfnamefont {Y.}~\bibnamefont {Chen}},\ }\bibfield  {title}
  {\bibinfo {title} {{Constraints on Low-Energy Effective Theories from Weak
  Cosmic Censorship}},\ }\href {https://doi.org/10.1103/PhysRevLett.126.031102}
  {\bibfield  {journal} {\bibinfo  {journal} {Phys. Rev. Lett.}\ }\textbf
  {\bibinfo {volume} {126}},\ \bibinfo {pages} {031102} (\bibinfo {year}
  {2021})},\ \bibinfo {note} {[Erratum: Phys.Rev.Lett. 126, 119903 (2021)]},\
  \Eprint {https://arxiv.org/abs/2006.08663} {arXiv:2006.08663 [gr-qc]}
  \BibitemShut {NoStop}%
\bibitem [{\citenamefont {Jiang}\ \emph {et~al.}(2021)\citenamefont {Jiang},
  \citenamefont {Sang},\ and\ \citenamefont {Zhang}}]{Jiang:2021ohh}%
  \BibitemOpen
  \bibfield  {author} {\bibinfo {author} {\bibfnamefont {J.}~\bibnamefont
  {Jiang}}, \bibinfo {author} {\bibfnamefont {A.}~\bibnamefont {Sang}},\ and\
  \bibinfo {author} {\bibfnamefont {M.}~\bibnamefont {Zhang}},\ }\bibfield
  {title} {\bibinfo {title} {{Comment on ''Constraints on Low-Energy Effective
  Theories from Weak Cosmic Censorship''}},\ }\href@noop {} {\  (\bibinfo
  {year} {2021})},\ \Eprint {https://arxiv.org/abs/2101.10172}
  {arXiv:2101.10172 [gr-qc]} \BibitemShut {NoStop}%
\bibitem [{\citenamefont {Lin}\ \emph {et~al.}(2023)\citenamefont {Lin},
  \citenamefont {Ning},\ and\ \citenamefont {Chen}}]{Lin:2022ndf}%
  \BibitemOpen
  \bibfield  {author} {\bibinfo {author} {\bibfnamefont {F.-L.}\ \bibnamefont
  {Lin}}, \bibinfo {author} {\bibfnamefont {B.}~\bibnamefont {Ning}},\ and\
  \bibinfo {author} {\bibfnamefont {Y.}~\bibnamefont {Chen}},\ }\bibfield
  {title} {\bibinfo {title} {{Weak cosmic censorship and the second law of
  black hole thermodynamics in higher derivative gravity}},\ }\href
  {https://doi.org/10.1103/PhysRevD.108.044025} {\bibfield  {journal} {\bibinfo
   {journal} {Phys. Rev. D}\ }\textbf {\bibinfo {volume} {108}},\ \bibinfo
  {pages} {044025} (\bibinfo {year} {2023})},\ \Eprint
  {https://arxiv.org/abs/2211.17225} {arXiv:2211.17225 [hep-th]} \BibitemShut
  {NoStop}%
\bibitem [{\citenamefont {Izumi}(2014)}]{Izumi:2014loa}%
  \BibitemOpen
  \bibfield  {author} {\bibinfo {author} {\bibfnamefont {K.}~\bibnamefont
  {Izumi}},\ }\bibfield  {title} {\bibinfo {title} {{Causal Structures in
  Gauss-Bonnet gravity}},\ }\href {https://doi.org/10.1103/PhysRevD.90.044037}
  {\bibfield  {journal} {\bibinfo  {journal} {Phys. Rev. D}\ }\textbf {\bibinfo
  {volume} {90}},\ \bibinfo {pages} {044037} (\bibinfo {year} {2014})},\
  \Eprint {https://arxiv.org/abs/1406.0677} {arXiv:1406.0677 [gr-qc]}
  \BibitemShut {NoStop}%
\bibitem [{\citenamefont {Reall}\ \emph {et~al.}(2014)\citenamefont {Reall},
  \citenamefont {Tanahashi},\ and\ \citenamefont {Way}}]{Reall:2014pwa}%
  \BibitemOpen
  \bibfield  {author} {\bibinfo {author} {\bibfnamefont {H.}~\bibnamefont
  {Reall}}, \bibinfo {author} {\bibfnamefont {N.}~\bibnamefont {Tanahashi}},\
  and\ \bibinfo {author} {\bibfnamefont {B.}~\bibnamefont {Way}},\ }\bibfield
  {title} {\bibinfo {title} {{Causality and Hyperbolicity of Lovelock
  Theories}},\ }\href {https://doi.org/10.1088/0264-9381/31/20/205005}
  {\bibfield  {journal} {\bibinfo  {journal} {Class. Quant. Grav.}\ }\textbf
  {\bibinfo {volume} {31}},\ \bibinfo {pages} {205005} (\bibinfo {year}
  {2014})},\ \Eprint {https://arxiv.org/abs/1406.3379} {arXiv:1406.3379
  [hep-th]} \BibitemShut {NoStop}%
\bibitem [{\citenamefont {Reall}(2021)}]{Reall:2021voz}%
  \BibitemOpen
  \bibfield  {author} {\bibinfo {author} {\bibfnamefont {H.~S.}\ \bibnamefont
  {Reall}},\ }\bibfield  {title} {\bibinfo {title} {{Causality in gravitational
  theories with second order equations of motion}},\ }\href
  {https://doi.org/10.1103/PhysRevD.103.084027} {\bibfield  {journal} {\bibinfo
   {journal} {Phys. Rev. D}\ }\textbf {\bibinfo {volume} {103}},\ \bibinfo
  {pages} {084027} (\bibinfo {year} {2021})},\ \Eprint
  {https://arxiv.org/abs/2101.11623} {arXiv:2101.11623 [gr-qc]} \BibitemShut
  {NoStop}%
\end{thebibliography}%

\end{document}